\pdfoutput=1

\documentclass[USenglish,oneside,twocolumn]{article}

\usepackage[utf8]{inputenc}
\usepackage[big]{dgruyter_NEW}
\usepackage{times}

\cclogo{\includegraphics{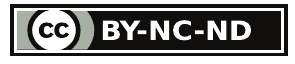}}

\usepackage{comment}
\usepackage{xspace}
\usepackage{color}
\usepackage{listings}
\usepackage{graphics, epstopdf, graphicx, epsfig, psfrag, epsf}
\usepackage{tabularx}
\usepackage{url}
\usepackage{breakurl}
\usepackage{booktabs}
\usepackage{xcolor,colortbl}
\usepackage{subcaption}
\usepackage{tikz}
\usepackage{calc}
\usepackage{multirow}
\usepackage{makecell}
\usepackage[normalem]{ulem} 

\usepackage{array}

\newcolumntype{P}[1]{>{\centering\arraybackslash}p{#1}}

\def\checkmark{\tikz\fill[scale=0.4](0,.35) -- (.25,0) -- (1,.7) -- (.25,.15) -- cycle;}
\def\scalecheck{\resizebox{\widthof{\checkmark}*\ratio{\widthof{x}}{\widthof{\normalsize x}}}{!}{\checkmark}}

\newcommand{\blue}[1]{\textcolor{blue}{#1}}
\newcommand{\red}[1]{\textcolor{red}{#1}}
\definecolor{lightgray}{rgb}{.9,.9,.9}
\definecolor{darkgray}{rgb}{.4,.4,.4}
\definecolor{purple}{rgb}{0.65, 0.12, 0.82}

\newcommand{\ie}{{\it i.e.}\xspace}
\newcommand{\eg}{{\it e.g.}\xspace}
\newcommand{\etc}{{\it etc.}\xspace}

\usepackage{pifont}
\newcommand{\xmark}{\ding{53}}%
\newcommand{\xmarkbold}{\ding{54}}%

\newcommand{\rokutool}{Rokustic}
\newcommand{\firetvtool}{Firetastic}
\newcommand{\roku}{Roku}
\newcommand{\firetv}{Fire~TV}
\newcommand{\testaccount}{SmartTV Test Account}
\newcommand{\smarttv}{smart TV}
\newcommand{\firetvtotalapps}{1010}
\newcommand{\rokutotalapps}{1044\xspace}
\newcommand{\comappcount}{128}
\newcommand{\droidbot}{DroidBot}
\newcommand{\descr}[1]{\smallskip\noindent\textbf{#1}}

\newcommand{\moab}{MoaAB\xspace}
\newcommand{\stopad}{SATV\xspace}
\newcommand{\firebog}{TF\xspace}
\newcommand{\piholelists}{PD\xspace}

\newcommand{\pihole}{Pi-hole\xspace}
\newcommand{\commapps}{common apps\xspace}

\newcommand{\ats}{ATS\xspace} 
\newcommand{\pii}{PII\xspace}

\newcommand{\rcs}{RCS}

\newcommand{\fqdn}{FQDN}

\newcommand{\esld}{eSLD}

\newcommand{\firstparty}{first party}
\newcommand{\thirdparty}{third party}

\DeclareFixedFont{\ttb}{T1}{txtt}{bx}{n}{8} 
\DeclareFixedFont{\ttm}{T1}{txtt}{m}{n}{8}  

\definecolor{deepblue}{rgb}{0,0,0.5}
\definecolor{deepred}{rgb}{0.6,0,0}
\definecolor{deepgreen}{rgb}{0,0.5,0}

\newcommand\pythonstyle{\lstset{
		language=Python,
		basicstyle=\ttm,
		otherkeywords={self},             
		keywordstyle=\ttb\color{deepblue},
		emph={MyClass,__init__},          
		emphstyle=\ttb\color{deepred},    
		stringstyle=\color{deepgreen},
		frame=tb,                         
		showstringspaces=false,
		commentstyle=\color{deepred}
}}

\lstnewenvironment{python}[1][]
{
	\pythonstyle
	\lstset{#1}
}
{}

\hypersetup{draft}

\begin{document}

  \author*[1]{Janus Varmarken}
	
  \author[2]{Hieu Le}
	
  \author[3]{Anastasia Shuba}
	
  \author[4]{Zubair Shafiq}
 
  \author[5]{Athina Markopoulou}

  \affil[1]{University of California, Irvine, E-mail: jvarmark@uci.edu}
	
  \affil[2]{University of California, Irvine, E-mail: hieul@uci.edu}
	
  \affil[3]{Broadcom Inc. (the author was a student at the University of California, Irvine at the the time the work was conducted), E-mail: ashuba22@gmail.com}

  \affil[4]{The University of Iowa, E-mail: zubair-shafiq@uiowa.edu}
	
  \affil[5]{University of California, Irvine, E-mail: athina@uci.edu}

  \title{\huge The TV is Smart and Full of Trackers}
  \subtitle{Towards Understanding the Smart TV Advertising and Tracking Ecosystem}
  \runningtitle{The TV is Smart and Full of Trackers}




  \begin{abstract}
{
Motivated by the growing popularity of smart TVs, we present a large-scale measurement study of smart TVs by collecting and analyzing their network traffic from two different vantage points.
First, we analyze aggregate network traffic of smart TVs in-the-wild, collected from residential gateways of tens of homes
%
and several different smart TV platforms, including Apple, Samsung, Roku, and Chromecast.
In addition to accessing video streaming and cloud services, we find that smart TVs frequently connect to well-known as well as platform-specific advertising and tracking services (\ats). 
Second, we instrument  \roku{} and Amazon \firetv{}, two popular smart TV platforms, by setting up a controlled testbed to systematically exercise the top-1000 apps on each platform, and analyze their network traffic at the granularity of the individual apps. 
%
We again find that smart TV apps connect to a wide range of \ats{}, and that the key players of the \ats{} ecosystems of the two platforms are different from each other and from that of the mobile platform.
%
%
Third, we evaluate the (in)effectiveness of state-of-the-art DNS-based blocklists in filtering advertising and tracking traffic for smart TVs.
We find that personally identifiable information (\pii{}) is exfiltrated to platform-related Internet endpoints and third parties, and that blocklists are generally better at preventing exposure of \pii{} to third parties than to platform-related endpoints. 
%
%
Our work demonstrates the segmentation of the \smarttv{} \ats{} ecosystem across platforms and its differences from the mobile \ats{} ecosystem, thus motivating the need for designing privacy-enhancing tools specifically for each \smarttv{} platform.
}
\end{abstract}

\maketitle

\section{Introduction}

Smart TV adoption has steadily grown over the last few years, with more than 37\% of US households with at least one smart TV  in 2018 which is a 16\% increase over 2017~\cite{smarttvhouseholdstats}.
The growth in smart TV is driven by two trends.
First, over-the top (OTT) video streaming services such as Netflix and Hulu have become quite popular, with more than 60 million and 28 million subscribers in the US, respectively~\cite{hulusubsstats}.
Second, smart TV solutions are available at relatively affordable prices, with many of the external smart TV boxes/sticks priced less than $\$50$ and built-in smart TVs priced on the order of hundreds of dollars \cite{smarttvprices}.
%
%
%
%
A diverse set of platforms, each with their own app store, are used by different smart TV products such as tvOS (Apple), Roku (Roku, TCL, Sharp), Android (Sony, Fire TV), SmartCast/Cast (Vizio, Chromecast), webOS (LG), etc.
%
%

The vast majority of smart TV apps are ad-supported \cite{adsupport}.
OTT advertising, which includes smart TV, is expected to increase by 40\% to \$2 billion in 2018 \cite{adforecast}.
Roku and Fire TV are two of the leading smart TV platforms in number of ad requests \cite{adgrowth}.
Despite their increasing popularity, the advertising and tracking services (``\ats'') on smart TV is currently not well understood by users, researchers, and regulators.
%
In this paper, we present a large-scale study of the smart TV advertising and tracking (\ats) ecosystem. %

\vspace{.05in} \noindent {\bf In the Wild Measurements (\S\ref{sec:in-the-wild}).}
First, we analyze the network traffic of smart TV devices in-the-wild.
We instrument residential gateways of 41 homes and collect flow-level summary logs of the network traffic generated by 57 smart TVs from 7 different platforms.
The comparative analysis of network traffic by different smart TV platforms uncovers  similarities and differences in their characteristics.
As expected, we find that a substantial fraction of the traffic is related to popular video streaming services such as Netflix and Hulu.
More importantly, we find generic as well as platform-specific advertising and tracking services.
Although realistic, the in-the-wild dataset does not provide app-level visibility, \ie we cannot determine which apps generate \ats  traffic. 
 To address this limitation, we undertake the following major effort.

\vspace{.05in} \noindent {\bf Controlled Testbed Measurements (\S\ref{sec:popular-app-testing}).}
We instrument and systematically study two popular smart TV platforms in a controlled environment.
Specifically, we design and implement \rokutool~for \roku{} and \firetvtool~for Amazon \firetv{} for systematically exercising apps and collecting their network traffic.
%
Using our tools, we analyze the top-1000 apps on Roku and Fire TV,  w.r.t. the top destinations contacted by those apps, at the granularity of fully qualified domain names (\fqdn{}s), effective second level domains (\esld{}s), and organizations.
We use ``domain (name)'' or ``endpoint'' interchangably in place of \fqdn{} and \esld{}. 
We further distinguish between first-party, third-party, and platform destinations, w.r.t. to the app that contacted them.
First, we find that certain platform-specific \ats{}, such as \url{logs.roku.com} for Roku and \url{amazon-adsystem.com} for Fire TV, are used by the vast majority of apps.
We also observe several third-party \ats, such as \url{doubleclick.net} and \url{scorecardresearch.com}, which are prevalent across many apps.
Second, we find that the same apps across different platforms (Roku vs. \firetv) have little overlap in terms of the endpoints they contact, which highlights the segmentation of smart TV platforms. 
Third,  we compare the smart TV platforms' \ats{} ecosystems to that of Android \cite{razaghpanah2018apps}, and identify  differences in terms of their key players.

\vspace{.05in} \noindent {\bf Evaluation of DNS-Based Blocklists (\S\ref{sec:eval-block-lists}).}
Since it is typically not viable to directly install ad/tracker blocking apps on smart TV platforms, 
smart TV users have to rely on DNS-based blocking solutions such as \pihole{} \cite{pihole-homepage}. 
We are interested in evaluating the effectiveness of well-known DNS-based blocklists in blocking advertising and tracking (\ats) domains accessed by different smart TV apps.
To that end,
 we examine and test four blocklists: (1) Pi-hole Default  blocklist (\textsf{\piholelists{}}) \cite{pihole-homepage}, (2) Firebog's recommended advertising and tracking lists  (\textsf{\firebog{}}) \cite{firebog}, (3) Mother of all Ad-Blocking (\textsf{\moab{}}) \cite{moab}, and (4) StopAd's  smart TV specific blocklist (\textsf{\stopad{}}) \cite{stopadtv}.
Our comparative analysis shows that block rates vary across different blocklists, with Firebog blocking the most and StopAd blocking the least.
We further evaluate and discuss the false negatives (FN) and false positives (FP).
We find that blocklists miss different \ats (FN), some of which are missed by all of the blocklists, while
more aggressive blocklists suffer from FP that result in breaking app functionality. 
We also identified many instances of \pii{} exfiltration, not only to \thirdparty{} \ats{}, but also to platform-specific endpoints, most of which seem unrelated to the core functionality of the app. Existing block lists seem more successful at blocking \pii{} exfiltration to third parties than to platform-specific endpoints.

\descr{Outline and Contributions.}
The structure of the rest of the paper is as follows. Section \ref{sec:related} provides background on smart TVs and reviews closely related work.
Section \ref{sec:in-the-wild} presents the  \emph{in-the-wild} measurement and analysis of 57 smart TVs from 7 different platforms in 41 homes.
Section \ref{sec:popular-app-testing} presents the {\em systematic} testing and analysis of the top-1000 Roku and Fire TV apps.
Section \ref{sec:eval-block-lists} evaluates four well-known {\em DNS-based blocklists} and show their limitations, including exfiltration of PIIs. 
Section \ref{sec:conclusion} concludes the paper and outlines directions for future work. The appendix provides additional graphs and details.
%
%

In summary, we perform a large-scale measurement study of the smart TV advertising and tracking (\ats) ecosystem.
Our work demonstrates the segmentation across smart TV platforms, limitations of current  DNS-based blocklists and   differences from mobile \ats{}.
This  motivates the need for further research on developing privacy-enhancing solutions specifically designed for each smart TV platform.
We plan to make our automated testing tools (\rokutool{} and \firetvtool{})  and collected datasets available to the community.

\section{Background \& Related Work \label{sec:related}}


\noindent {\bf Background on Smart TVs.}
A smart TV is essentially an Internet-connected TV that has apps for users to stream content, play games, and even browse the web.
There are two types of smart TV products in the market: (1) built-in smart TVs, and (2) external smart TV box/stick.
On one hand, TV manufacturers such as Samsung and Sony now offer TVs with built-in smart TV functionality.
On the other hand, several external box/stick solutions such as Roku (by Roku), Fire TV (by Amazon), Apple TV (by Apple), and Chromecast (by Google) are available to convert a regular TV into a smart TV.
Some TV manufacturers have started to integrate external box/stick solutions into their smart TVs.
For example, TCL and Sharp offer smart TVs that integrate Roku TV while Insignia and Toshiba offer \firetv{} instead within their \smarttv{s}.

There is a diverse set of smart TV platforms, each with its own set of apps that users can install on their TVs.
Many smart TVs use an Android-based operating system (\eg Sony, AirTV, Philips) or a modified version of it (\eg Fire TV).
Regular Android TVs have access to apps from the Google Play Store, while Fire TV has its own app store controlled by Amazon.
In both cases, applications for such TVs are built in a manner similar to regular Android applications.
%
%
Likewise, Apple TV apps are built using technologies and frameworks that are also available for iOS apps, and both types of apps can be downloaded from Apple's App Store.

Some smart TV platforms are fairly distinct as compared to traditional Android or iOS.
For example, Samsung smart TV apps are built for their own custom platform called Tizen and are downloadable from the Tizen app store.
Likewise, applications for the Roku platform are built using a customized language called BrightScript, and are accessible via the Roku channel store.
Yet another line of smart TVs such as LG smart TV and Hybrid broadcast broadband TV (HbbTV) follow a web-based ecosystem where applications are developed using HTML, CSS, and JavaScript.
Finally, some smart TV platforms do not have app stores of their own, but are only meant to ``cast'' content from other devices such as smartphones.
For instance, Chromecast provides users the ability to stream content from their mobile device or laptop to their TV.

As with mobile apps, smart TV apps can integrate third-party libraries and services, often for advertising and tracking purposes.
Serving advertisements is one of the main ways for smart TV platforms and app developers to generate revenue \cite{adsupport}.
Roku and Fire TV are two of the leading ad-supported smart TV platforms \cite{adgrowth}.
Roku's advertising revenue exceeded \$250 million in 2018 and is expected to more than double by 2020 \cite{rokuadsgrowing}.
Both Roku and Fire TV take 30\% cut of the advertising revenue from apps on their platforms \cite{adsalesfiretv}.
Smart TV advertising ecosystem mirrors many aspects of the vanilla web advertising ecosystem.
Most importantly, smart TV advertising uses programmatic mechanisms that allow apps to sell their ad inventory in an automated fashion using behavioral targeting \cite{rokuadsolutions,firetvadsolutions}.

The rapidly growing smart TV advertising and associated tracking ecosystem has already warranted privacy and security investigations into different smart TV platforms.
Consumer Reports examined privacy policies of various smart TV platforms including Roku, LG, Sony, and Vizio \cite{consumerreporttv}.
They found that privacy policies are often challenging to understand and it is difficult for users to opt out of different types of tracking.
For instance, many smart TVs use Automatic Content Recognition (ACR) to track their users' viewing data and use it to serve targeted ads \cite{nytimestvtracking}.
Vizio paid \$2.2 million to settle the charges by the Federal Trade Commission (FTC) that they were using ACR to track users' viewing data without their knowledge or consent \cite{ftcvizio}.
While smart TV platforms now allow users to opt-out of such tracking, it is not straightforward for users to turn it off \cite{nytimesstoptracking}.
Further, even with ACR turned off, users still must agree to a basic privacy policy that asks for the right to collect data about users' location, choice of apps, \etc

\descr{Related Work.}
While the desktop~\cite{mayer2012,gill2013followthemoney,englehardt2016} and mobile~\cite{razaghpanah2018apps,recon,shuba2018nomoads} \ats{} ecosystems have been thoroughly studied, the smart TV \ats{} ecosystem has not been examined at scale until recently.
Concurrently with our work, three other papers also studied the network behavior and privacy implications of \smarttv{}s~\cite{ren2019, huang2019, moghaddam2019}.
Ren et al.~\cite{ren2019} studied a large set of IoT devices, spanning multiple device categories. 
Their results showed that \smarttv{}s were the category of devices that contacted the largest number of third parties, which further motivates our in-depth study of the \smarttv{} \ats{} ecosystem.
Huang et al.~\cite{huang2019} used crowdsourcing to collect network traffic for IoT devices in the wild and showed that \smarttv{}s contact many trackers by matching the contacted domains against the Disconnect blocklist.
Finally, and most closely related to our work, Moghaddam et al. \cite{moghaddam2019} also instrumented the \roku{} and \firetv{} platforms to map the \ats{} endpoints contacted by the top-1000 apps of each platform, as well as the exposure of PII.
Our work combines the individual merits of these concurrent works, providing a holistic view of the \smarttv{} \ats{} ecosystem by analyzing a total of eight different \smarttv{} platforms in-the-wild and in-depth in the lab.
We further contribute to this existing literature along two fronts: 
First, we show that even the same app across different \smarttv{} platforms contact different \ats{}, which shows the fragmentation of the \smarttv{} \ats{} ecosystem.
Second, we evaluate the effectiveness of different sets of blocklists, including a \smarttv{} specific blocklist, in terms of their ability to prevent ads as well as their adverse effects on app functionality. In addition to blocklist evaluation, we also suggest ways to aid blocklist curation through analysis of app prevalence and PII exposures.

%
%
Earlier work in this space include that of Ghiglieri and Tews~\cite{ghiglieri2014}, who studied how broadcasting stations could track viewing behavior of users in the HbbTV platform.
In contrast to the rich app-based platforms we study, the HbbTV platform studied in~\cite{ghiglieri2014} contained only one HbbTV app that uses HTML5-based overlays to provide interactive content.
Related to our work, they found that the HbbTV app loaded third-party tracking scripts from Google Analytics.
Malkin et al. \cite{malkin2018can} surveyed 591 U.S. Internet users about their expectations on data collection by smart TVs.
They found that users would rather enjoy new technology than worry about privacy, and users thus over rely on existing laws and regulations to protect their data.

\section{Smart TV Traffic in the Wild}
\label{sec:in-the-wild}

\begin{table}[b!]
	\small
	\begin{center}
		\label{tab:devices}
		\begin{tabular}{c|c|c|c|c}
			\textbf{Smart} & \textbf{Device} &  \textbf{Average } & \textbf{Average }  & \textbf{Average }\\
			\textbf{TV} & \textbf{Count} &  \textbf{Flow} & \textbf{Flow}  & \textbf{\esld{}}\\
			\textbf{Platform} & \textbf{} &  \textbf{Count} & \textbf{Volume}  & \textbf{Count}\\
			\textbf{} & \textbf{} &  \textbf{/Device} & \textbf{/Device} & \textbf{/Device}\\
			\textbf{} & \textbf{} &  \textbf{(x 1000)} & \textbf{(GB)} & \\
			
			\hline
			Apple & 16 & 49.3 & 46.6 & 536\\
			Samsung & 11 & 62.6 & 33.2 & 369 \\
			Chromecast & 10 & 201.9 & 26.3 & 354 \\
			Roku & 9 & 48.1& 83.0 & 543\\
			Vizio & 6 & 43.4 & 63.4 & 278\\
			LG & 4 & 10.9 & 0.9 & 189\\
			Sony & 1 & 33.1 & 0.1 & 186 \\
		\end{tabular}
	\end{center}
	\caption{Traffic statistics of 57 smart TV devices observed across 41 homes (``in-the-wild'' dataset).}
	\label{tbl:in-the-wild-stats}
\end{table}

\begin{figure*}[t!]
	
	\begin{subfigure}{1\columnwidth}
		\centering
		\includegraphics[width=0.85\linewidth]{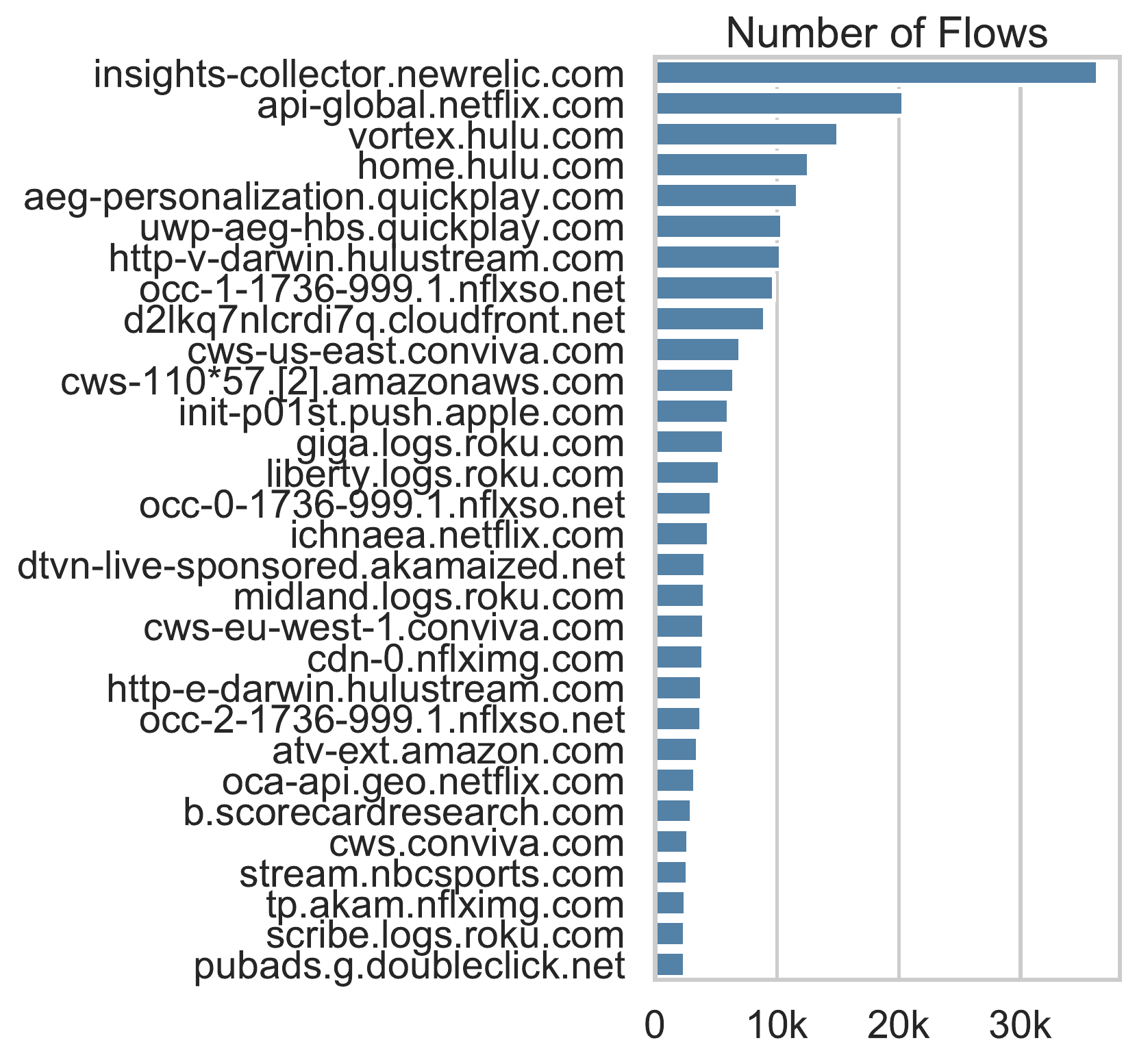}
		\caption{\roku{}}
		\vspace{-5pt}
	\end{subfigure}
	\hspace{0.6cm}
	\begin{subfigure}{1\columnwidth}
		\centering
		\includegraphics[width=0.85\linewidth]{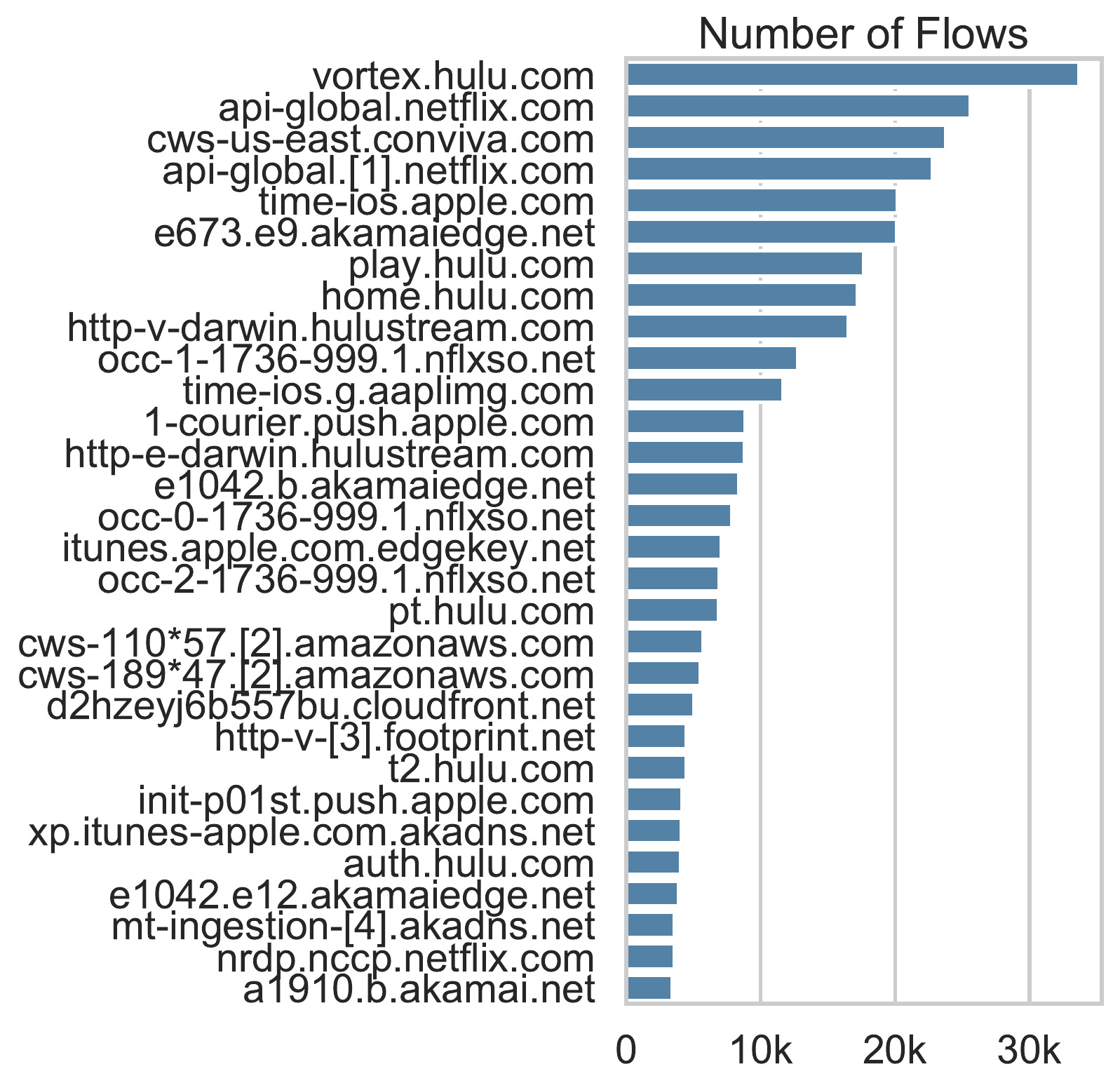}
		\caption{Apple}
	\end{subfigure}
	\begin{subfigure}{1\columnwidth}
		\centering
		\includegraphics[width=0.8\linewidth]{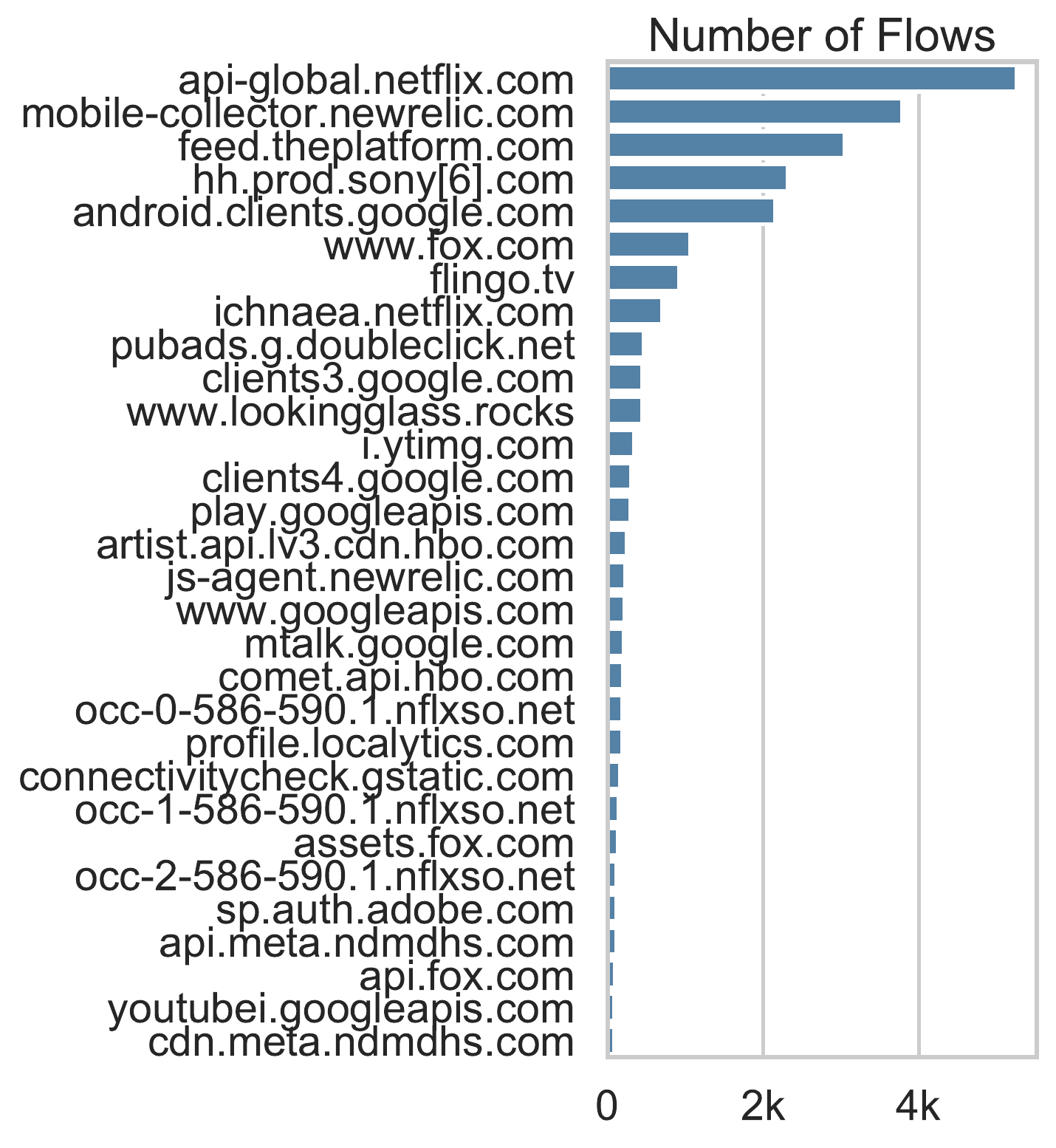}
		\caption{Sony}
		\vspace{-5pt}
	\end{subfigure}
	\hspace{0.6cm}
	\begin{subfigure}{1\columnwidth}
		\centering
		\includegraphics[width=0.87\linewidth]{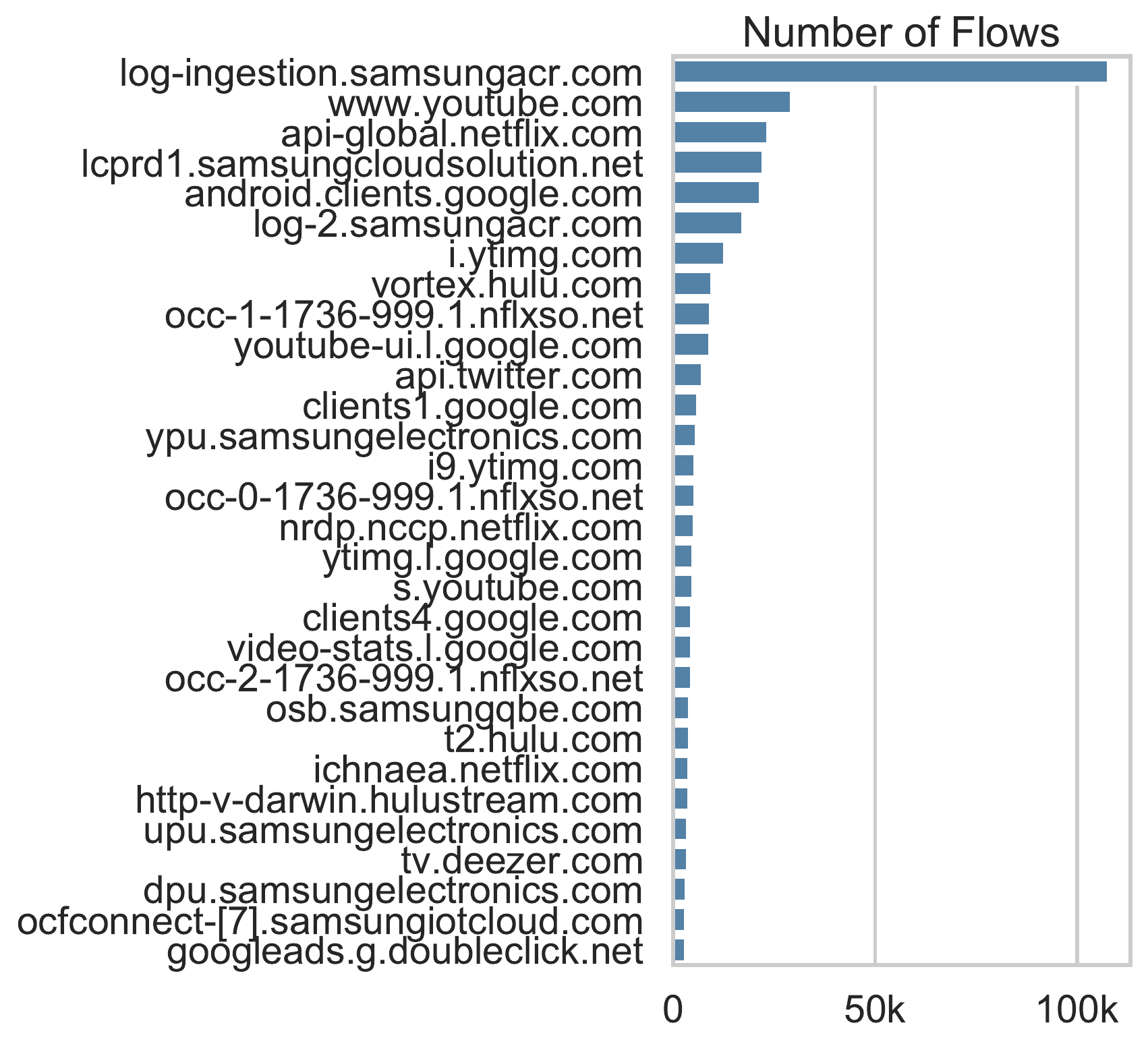}
		\caption{Samsung}
		\vspace{-5pt}
	\end{subfigure}
	\caption{Top-30 fully qualified domain names in terms of number of flows per device for a subset of the smart TVs in the ``in-the-wild'' dataset. See Appendix~\ref{sec:inthewild-complete-appendix} for the other brands.}
	\label{fig:in-the-wild-fig}
\end{figure*}

\vspace{.05in} \noindent \textbf{Data Collection.}
To study smart TV traffic characteristics in the wild, we monitor network traffic of 41 homes in a major metropolitan area in the United States.
We sniff network traffic of smart TV devices at the residential gateways using off-the-shelf OpenWRT-capable commodity routers.
We collect flow-level summary information for network traffic.
For each flow, we collect its start time, \fqdn{} of the external endpoint (using DNS), and the internal device identifier.
We identify smart TVs using heuristics that rely on DNS, DHCP, and SSDP traffic and also manually verify the identified smart TVs by contacting users.
Our data collection covers a total of 57 smart TVs across 41 homes over the duration of approximately 3 weeks in 2018.
Note that we obtained written consent from users, informing them of our data collection and research objectives, in accordance with our institution's IRB guidelines.

\vspace{.05in} \noindent \textbf{Dataset Statistics.}
Table \ref{tbl:in-the-wild-stats} lists basic statistics of smart TV devices observed  in our dataset.
Overall, we note 57 smart TVs from 7 different vendors/platforms using a variety of technologies such as Apple TV (tvOS), Samsung Smart TV (Tizen), Google Chromecast (Cast SDK), Roku (standalone TV and HDMI sticks), Vizio (SmartCast based on Chromecast), LG Smart TV (webOS), and Sony Smart TV (Android).

These smart TV devices account for substantial traffic both in terms of number and volume of traffic flows.
We indeed expect smart TV devices to generate significant traffic because they are typically used for OTT video streaming \cite{erman2011over}.
First, we note that all smart TV devices generate tens of thousands of traffic flows on average.
Chromecast devices generate the highest number of flows (exceeding 200 thousand flows) on average.
Samsung, Apple, Roku, and Vizio devices generate nearly 50 thousand flows on average.
Second, we note a similar trend for average flow volume.
Roku devices generate the highest volume of flow (exceeding 80 GB) on average, with one Roku device generating as much as 283 GB.
Except for LG and Sony devices, all smart TV devices generate at least tens of GBs worth of traffic on average.
Finally, we also note that smart TV devices typically connect to hundreds of different endpoints on average.
As we discuss next, smart TV devices connect to a variety of endpoints for video streaming and related functionality.


\vspace{.05in} \noindent \textbf{Endpoint Analysis.}
Fig. \ref{fig:in-the-wild-fig} plots the top-30 \fqdn{}s in terms of flow count for Roku, Apple, Sony, and Samsung smart TV platforms.
The plots for the remaining smart TV platforms are in Appendix \ref{sec:inthewild-complete-appendix}.
We notice several similarities in the domains accessed by different smart TV devices.
First, as expected, we note that popular video streaming services such as Netflix and Hulu are popular across the board.
For example, domains such as \url{api-global.netflix.com} and \url{vortex.hulu.com} appear for different smart TV platforms.
Second, we note that cloud/CDN services such as Akamai and AWS (Amazon) also appear for different smart TV platforms.
For example, domains such as \url{*.akamai*.net} and \url{*.amazonaws.com} appear across most smart TVs.
Smart TVs likely connect to cloud/CDN services because popular video streaming services such as Netlifx typically rely on \thirdparty{} CDNs \cite{bottger2018open,adhikari2012tale}.
Third, we note the prevalence of well-known advertising and tracking services (\ats{}).
For example, \url{*.scorecardresearch.com} and \url{*.newrelic.com} 
are well-known third-party tracking services, and \url{pubads.g.doubleclick.net} is a well-known third-party advertising service.

We notice several platform-specific differences in the domains accessed by different smart TV platforms.
For example, \url{giga.logs.roku.com} (Roku), \url{time-ios.apple.com} (Apple), \url{hh.prod.sonyentertainmentnetwork.com} (Sony), and \url{log-ingestion.samsungacr.com} (Samsung) are unique to different types of smart TVs.
In addition, we notice platform-specific \ats{}.
For example, the following advertising-related domains are not in the top-30 (and therefore not pictured in Fig. \ref{fig:in-the-wild-fig}), but are unique to different smart TV platforms: \url{p.ads.roku.com} (Roku), \url{ads.samsungads.com} (Samsung), and \url{us.info.lgsmartad.com} (LG).

\begin{figure}[t!]
	\vspace{-5pt}
	\centering
	\includegraphics[width=0.73\columnwidth]{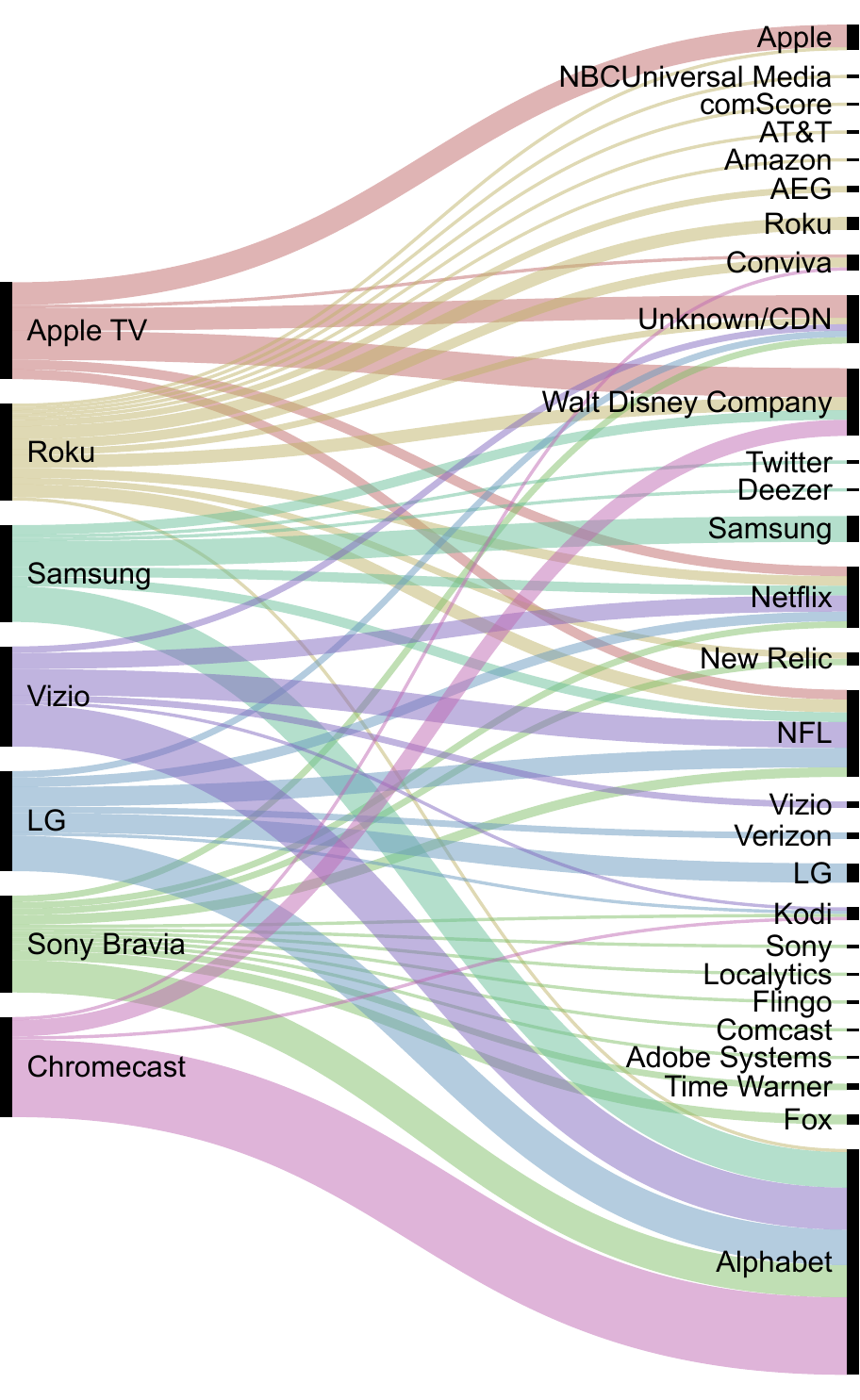}
\vspace{-.1in}
	\caption{Mapping of platforms measured in-the-wild to the ultimate parent organizations of the endpoints they contact  (for the top-30 \fqdn{}s of each platform). The width of an edge indicates the number of distinct \fqdn{}s within that organization that was accessd by the platform.  
	}
	\label{fig:inthewild-flow-all}
\vspace{-.2in}
\end{figure}

\vspace{.05in} \noindent \textbf{Organizational Analysis.} To understand the role of different parent organizations within each smart TV platform, we map each \fqdn{} to its effective second level domain (\esld{}) using Mozilla's Public Suffix List \cite{mozilla-public-suffix-list,tldextract}.
We then use Crunchbase \cite{crunchbase-homepage} to find the organization name and follow Crunchbase's acquisition and sub-organization information to find the ultimate parent company.
For example, \url{hulu.com} belongs to the Walt Disney Company and \url{youtube.com} belongs to Alphabet.
Figure~\ref{fig:inthewild-flow-all} illustrates the mix of different parent organizations contacted by the seven \smarttv{} platforms in our dataset.
The illustration shows the prevalence of Alphabet in smart TV platforms like Chromecast, Sony, Samsung, LG, and Vizio, while also revealing competing organizations such as Apple on the other end of the spectrum.
The illustration also shows that Roku and Sony contact a wide range of \ats{} organizations like Conviva, Localytics, and Adobe Services.

\vspace{.05in} \noindent \textbf{Takeaway \& Limitations.}
Traffic analysis of different smart TV platforms in the wild highlights interesting similarities and differences.
As expected, all devices generate  traffic related to popular video streaming services.
In addition, they also access advertising and tracking services (\ats), both well-known and platform-specific.
While our vantage point at the residential gateway provides a real-world view of how smart TV devices behavior, it lacks granular information beyond flows (\eg{}, packet-level information) and the specific apps that generate the traffic.
%
%
Another limitation of in the wild analysis is that our findings may be biased by the viewing habits of users in these 41 households.
It is not clear how to normalize our analysis to provide a fair comparison of endpoints accessed by different smart TV platforms.
We address these limitations next by systematically analyzing two popular smart TV platforms in a controlled testbed.

\section{Systematic Testing of the \roku{} and \firetv{} Platforms}
\label{sec:popular-app-testing}
In this section, we focus on two of the popular smart TV platforms, namely \roku{} and Amazon \firetv{}.
Sections \ref{sec:roku} and \ref{sec:firetv} present our measurement approach for systematically testing approximately 1000 apps in each platform while collecting their network traffic\footnote{Since testing is done automatically, no real users are involved, thus no IRB is needed.}.
Our measurement approaches enables us to automatically test a large number of apps.  More importantly, it provides visibility into the behaviors of individual apps, which was not possible from the vantage point used for the in-the-wild dataset in the previous section.
In Sections \ref{sec:comparison} and \ref{sec:eco-compare-common-apps}, we summarize and analyze the two datasets, and compare them to each other as well as to the in-the-wild datasets of the previous section, and to the Android \ats{} ecosystem \cite{razaghpanah2018apps}.

\subsection{Roku Data Collection \label{sec:roku}}
In this section, we first present an overview of the Roku platform.
Next, we present our app selection method.
We then proceed to describe \rokutool{}---the software tool we wrote for automatically exercising Roku apps.
The data resulting from exercising \rokutotalapps{} \roku{} apps provides insight into what \ats{} are most prevalent on the Roku platform.

\descr{Roku Platform.}
We start by describing the Roku TV (referred to as Roku for the rest of the paper) platform, which has its own app store that offers more than 8500 apps, called ``channels''.
We will use the terms ``apps'' and ``channels'' interchangeably for the rest of the paper.
For security purposes, Roku sandboxes each app (apps are not allowed to interact or access the data of other apps) and provides limited access to system resources~\cite{rokusecurity}.
Furthermore, Roku apps cannot run in the background.
Specifically, app scripts are only executed when the user selects a particular app, and when the user exits, the script is halted, and the system resumes control~\cite{rokubackgroundapps}.


To display ads, apps typically rely on the Roku Advertising Framework which is integrated into the Roku SDK~\cite{rokuadsframework}.
The framework allows developers to use ad servers of their preference and updates automatically without requiring the developer to rebuild the app.
Even though such a framework eliminates the need for \thirdparty{} \ats libraries, the development and usage of such libraries is still possible.
For example, the Ooyala IQ SDK \cite{ooyalasdk} provides various analytics services that can be integrated into a Roku app.
Thus, although the Roku sandboxes apps, such libraries can help \ats servers learn the viewing habits of users by collecting data from multiple apps.
In terms of permissions, Roku only protects microphone access with a permission and does not require any permission to access the advertising ID.
Users can choose to reset this ID and opt-out of targeted advertising at any time~\cite{rokuadsframework}.
However, apps and libraries can easily create other IDs or use fingerprinting techniques to continue tracking users even after opt-out.

\descr{App Selection.}
The Roku Channel Store \cite{rokuchannelstore} (\rcs{}) provides a web interface for browsing the set of available Roku apps, and allows for ``one-click'' installation of Roku apps on Roku devices linked to a Roku user account.
To the best of our knowledge, Roku does not provide any public documentation on how to query the \rcs{} in a programmatic way.
We therefore reverse-engineered the REST API that provides the data to the \rcs{} web interface by inspecting the HTTP(S) requests sent by the browser while manually browsing the \rcs{}.
Using this insight, we wrote a script that crawls the \rcs{} for the metadata of all Roku apps.
The script first issues an API call to fetch all app categories.
Next, for each category, it performs a series of API calls to determine the app IDs of all apps in that category.
Finally, for each app ID, an API call is made to fetch the full metadata for that app.
At the time of writing, this resulted in a total of 8,515 different Roku apps.

To ensure that we test the most relevant apps, we selected the top 50 apps in 30 out of the total 32 categories.
We excluded ``Themes'' and ``Screensavers'' since these apps do not show up among the regular apps on the Roku device and therefore cannot be operated using our automation software described below.
Since statistics for the number of downloads or installs are not provided by Roku, we based our selection on the ``star rating count''---which we interpret as the review count---present in the metadata extracted from the \rcs{}.
Note that Roku apps can be labeled with multiple categories, meaning some apps contribute to the top 50 of multiple categories.
Furthermore, some categories contain fewer than 50 apps, a handful of apps could not be installed due to incompatibility with the Roku Express (Roku offers a range of more capable devices), and about a dozen of apps had to be discarded due to failure during automation.
This places the final count of apps in our dataset at \rokutotalapps{}.

\descr{Automation (\rokutool).}
To scale testing of apps, we implement a software system, \rokutool{}, that automatically installs and exercises a given set of Roku apps.

{\em Setup and Network Traffic capture.}
We run \rokutool{} on a Raspberry Pi 3 Model B+ set up to host a standalone network as per the instructions given in \cite{raspberrypiap}.
The Pi's \texttt{wlan0} interface is configured as a wireless access point with DHCP server and NAT, and the Roku Express is connected to this local wireless network.
The Pi's \texttt{eth0} interface connects the Pi and the Roku Express to the WAN.
This setup enables us to collect all traffic going in and out of the Roku Express by running \texttt{tcpdump} on the Raspberry Pi's \texttt{wlan0} interface.

{\em App Exploration.}
From manual inspection of a few apps (\eg YouTube and Pluto TV), we found that playable content is often presented in a grid, where each cell is a different video or live TV channel.
Generally, the user interface defaults to highlighting one of these  cells (\eg{}, the first recommended video).
Pressing ``SELECT'' on the Roku remote immediately after the app has launched will therefore result in playback of some content.
From this insight, we devised a simple algorithm that attempts to cause playback of three different videos for each installed Roku app.  Due to lack of space, we only provide an overview below, but we plan to make the tool publicly available.
 The algorithm utilizes the Roku External Control Protocol (ECP) API \cite{rokuecp} to mimic a user's interaction with each Roku app.
The ECP turns the Roku into an HTTP-server with a REST-like API.
The API includes an endpoint for each key on the physical Roku remote.
This allows us to send virtual key presses to the Roku.
We combine this with the ECP endpoint that allows for querying the Roku device for its set of installed apps, as well as the endpoint that launches a specific app, to cycle through and exercise all installed apps.

{\em Putting it all together.} For each Roku app, the algorithm first starts a packet capture so as to produce a \texttt{.pcap} file for each Roku app, thereby essentially labeling traffic with the app that caused it.
Since Roku does not allow apps to execute in the background (see ``Roku Platform'' earlier in this section), all traffic captured during execution of a single app will belong to that app and the Roku system.
The target app is then launched, and the algorithm pauses, waiting for the app to load.
A virtual ``SELECT'' key press is  then sent to start video playback, and the algorithm subsequently pauses for five minutes to let the content play.
The app is then relaunched by returning to the Roku's home screen and then launching the app again. 
A different video/live TV channel is then selected by sending a sequence of navigational key presses followed by a ``SELECT'' key press, and the algorithm waits another five minutes for the content to play.
We perform two such relaunches, making the total interaction time with each app approximately 16 minutes (due to sleep timers).

\subsection{\firetv{} Data Collection \label{sec:firetv}}
In this section, we first briefly describe the \firetv{} platform.
We then describe \firetvtool{}---our automated methodology for systematically testing and collecting traffic from \firetvtotalapps{} \firetv{} apps.
The dataset reveals properties of \ats{} for \firetv{} devices and its analysis is deferred to Sections~\ref{sec:eco-compare} and \ref{sec:eco-compare-common-apps}.

\descr{\firetv{} Platform.}
Although \firetv{} is made by Amazon, its underlying operating system, Fire OS, is a modified version of Android.
This allows apps for \firetv{} to be developed in a similar fashion to Android apps.
Therefore, all third-party libraries that are available for Android apps can also be integrated into \firetv{} apps.
Similarly, application sandboxing and permissions in \firetv{} are analogous to that of Android, and any permission requested by the app is inherited by all libraries that the app includes.
This allows third-party libraries to track users across apps using a variety of identifiers, such as Advertising ID, Serial Number, Device ID, and Account  Names, etc. We further discuss tracking through PII exposure in Section~\ref{sec:pii-exposure}.

\descr{App Selection.}
Amazon's app store promotes apps with a curated list of Top Featured apps. 
Our data collection depends on this list to determine the relevant applications to test for \firetv{}. 
Using the store's web interface, which allows installations of apps in one click, we manually install \firetvtotalapps{} apps in batches of 50. 
Amazon's app store had around 4,000 free apps at the time of writing, thus our dataset covers approximately 25\%. 
Along the way, we ignore some apps that use a local VPN, that could not be installed manually, and utility apps that can change the device settings (which would affect the test environment). 
As a result, we ignored around 200 apps while collecting \firetvtotalapps{} testable applications.

\descr{Automation (\firetvtool).}
To scale testing of \firetv{} apps, we build a software tool, \firetvtool{}, that integrates the capabilities of two open source tools for Android: an SDK for network traffic collection~\cite{shuba2016, antmonitoross} and a tool for input automation~\cite{li2017droidbot}.  

{\em On-Device Network Traffic Collection.} Since \firetv{} is based on Android, we can use existing Android tools to capture network traffic. Although there are various methods for capturing traffic on Android on the smartTV device itself (\eg \texttt{androidtcpdump} \cite{androidtcpdump}), most of them require a rooted device. While it is possible to root a \firetv{}, it may make applications behave differently if they detect root. Thus, to collect measurements that are representative of an average user, we use a VPN-based traffic interception method that does not require rooting the device. Specifically, we used an open-source VPN-based library \cite{shuba2016, antmonitoross} to intercept all incoming and outgoing network traffic (including decrypting TLS connections) from the \firetv{} while labeling each packet with the package name of the application that generated it. In addition, we monitor network traffic from system applications  to learn how the device behaves with each different application.

{\em App Exploration.} To automatically explore each \firetv{} application, we utilize~\cite{li2017droidbot}, a Python application that sends commands to an Android device via the Android Debug Bridge (ADB) to simulate inputs such as pressing buttons, input text in a search box, or selecting a video to play.  It does not require rooting the device and is easily customizable. This tool is ideal for testing \firetv{} applications because it treats each application as a tree of possible paths to explore instead of randomly generating events. We configured it to utilize its breadth first search algorithm to explore each application, and ensure various videos and ads would be played.
Furthermore, we deduce that developers would minimize the necessary clicks in order to reach the core sections of their applications, especially for playing video content. With some trial and error, we selected the input command interval as three seconds which leaves enough time for applications to handle the command and load the next view during app exploration.

{\em Putting it all together.} In summary, \firetvtool{} does the following for each app:  (1) it starts the local VPN, (2) explores the app for 15 minutes, (3) stops the local VPN, and (4) extracts the \texttt{.pcapng} files that were generated during testing. We use \firetvtool{} to do testing in parallel with six \firetv{} devices.  Our test setup is resource-efficient and scalable: we use only one computer to send commands to multiple (6 in our experiments)  \firetv{} devices, and we are able to collect network traffic data for \firetvtotalapps{} apps within a one week period.

\subsection{Comparing \roku{} and \firetv{}\label{sec:comparison}}
\label{sec:eco-compare}

\descr{Overview.}
We start by summarizing the datasets collected using \rokutool{} and \firetvtool{} in Table~\ref{tab:roku-firetv-dataset}.
Since these datasets provide app-level granularity, we report the number of \fqdn{}s and URL paths that are contacted by one or multiple apps.
For \roku{}, we find 2191 distinct \fqdn{}s, 699 of which are contacted by more than one app. 
For \firetv{}, we find 1734 distinct \fqdn{}s, 603 of which are contacted by multiple apps. 
We also find 578 \fqdn{}s that appear in both datasets, 199 of which are contacted by more than one app.

\descr{Categorizing Destinations.}
The app-level visibility also enables us to categorize Internet destinations where packets are sent to as first party, third party, or platform party, by comparing that destination with the app that generated the packet.
To this end, we first map \fqdn{}s to \esld{}s using Mozilla's Public Suffix List \cite{mozilla-public-suffix-list,tldextract} and then categorize each \esld{} as \firstparty{}, \thirdparty{}, or platform party as follows:

\begin{enumerate}
\item We first tokenize app identifiers and \esld{}s. 
For \firetv{}, we tokenize the package names while relying on app and developer names for \roku{} since its apps do not have package names.

\item We then match the app's tokenized identifier against the tokenized \esld{}. 
In doing so, we ignore common tokens or platform-specific strings like ``com'', ``firetv'', ``roku'' ``free'', ``paid'' etc. 
If the tokens match, we label the \esld{} as \emph{\firstparty{}}. 
If the tokens do not match, and if the \esld{} is contacted by at least two different apps (from different developers), we label it as \emph{\thirdparty{}}.

\item Finally, we label an \esld{} as \emph{platform-specific party} if it originated from platform activity rather than app activity. 
For \firetv{}, the VPN Tool \cite{shuba2016} labels connections with the responsible process. 
For \roku{}, we simply check if the \esld{} contains the string ``roku.''

\end{enumerate}

\begin{figure}[t!]
	\centering
	\begin{subfigure}{0.9\columnwidth}
		\centering
		\includegraphics[width=0.95\columnwidth]{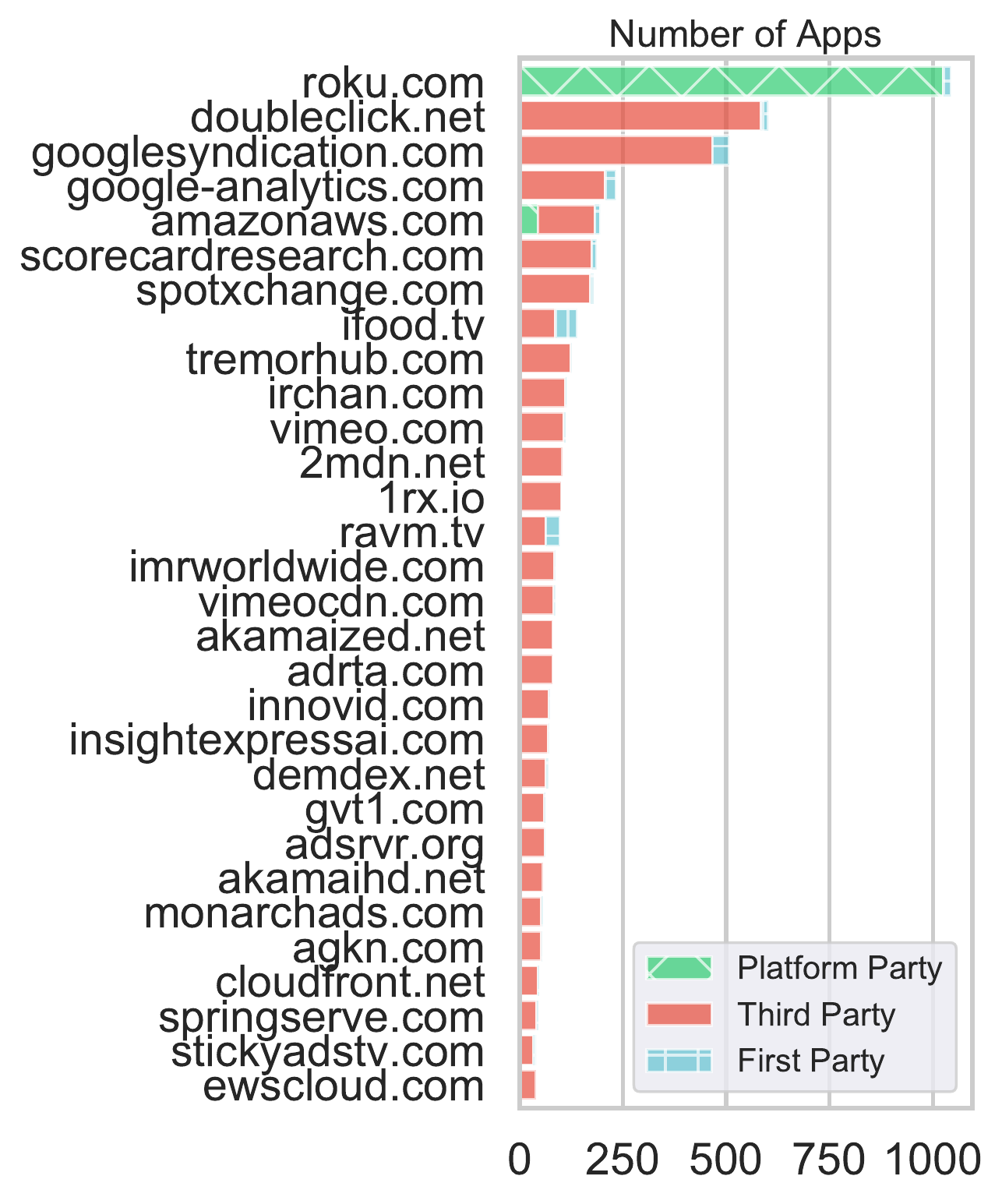}
		\caption{App penetration for the top-30 eSLDs in the \roku{} testbed dataset.}
		\label{fig:roku-top-hostnames-by-channels-sld}
	\end{subfigure}\hfill%
	\begin{subfigure}{0.9\columnwidth}
		\centering
		\includegraphics[width=0.78\columnwidth]{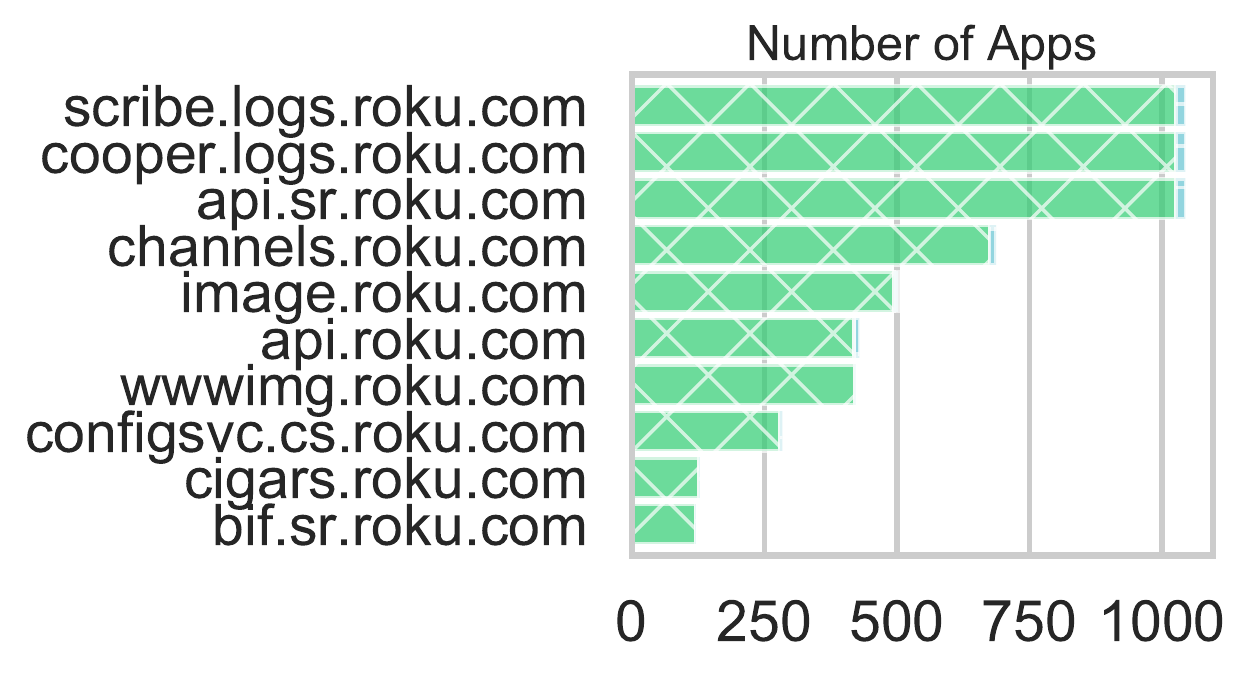}
		\caption{\roku{} top-10 platform-specific domains.}
		\label{fig:roku-top-hostnames-by-channels}
	\end{subfigure}\hfill%
	\begin{subfigure}{0.9\columnwidth}
		\centering
		\includegraphics[width=0.95\columnwidth]{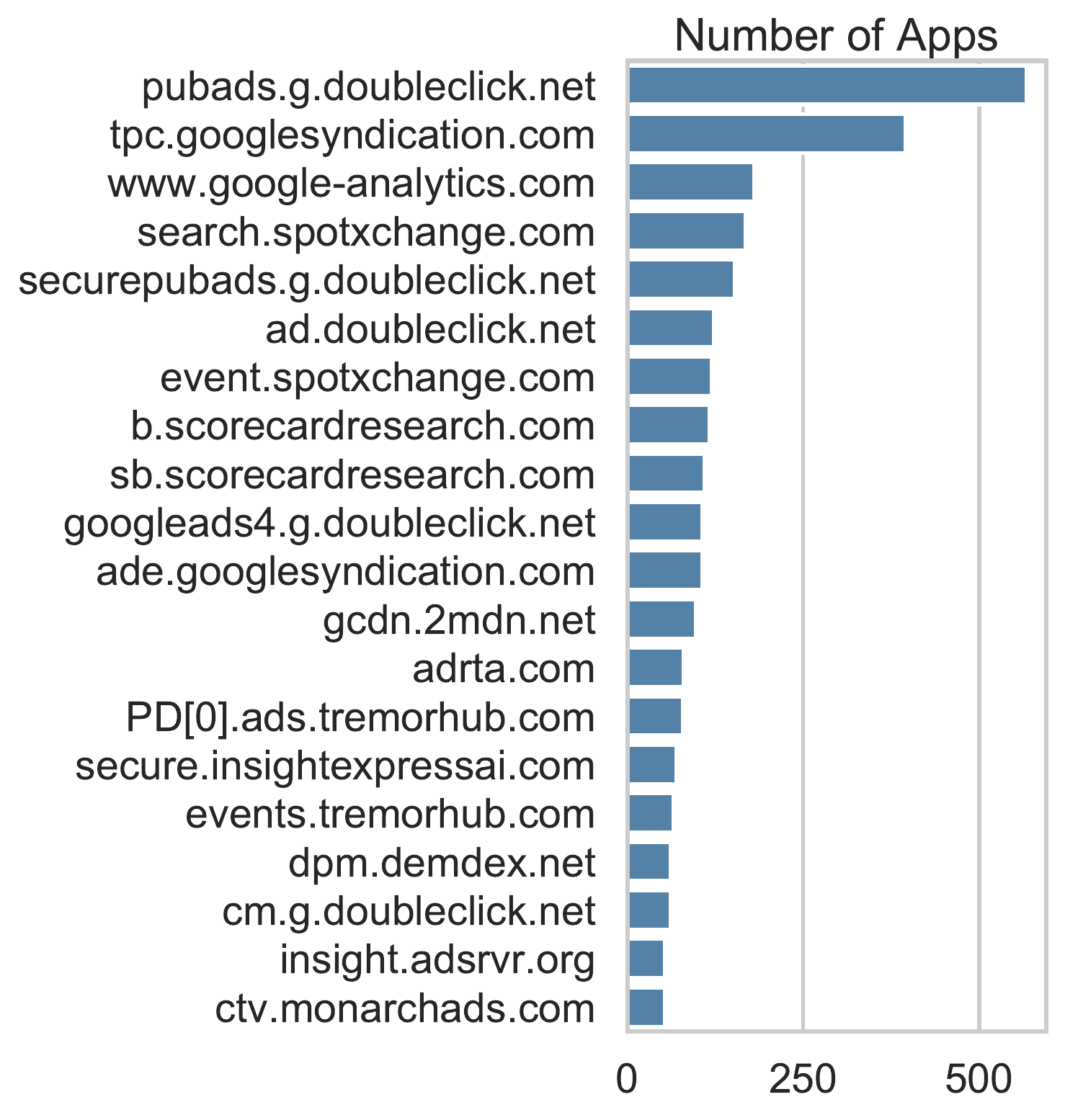}
		\caption{\roku{} top-20 third party \ats{} domains.}
		\label{fig:roku-top-thirdparty-ats}
	\end{subfigure}\hfill%
	\caption{\roku{} testbed dataset broken down by top eSLDs, platform-specific domains, and third party \ats{} domains.}
	\label{fig:roko-eco-all}
\end{figure}

\begin{figure}[t!]
	\centering
	\begin{subfigure}{0.9\columnwidth}
		\centering
		\includegraphics[width=0.96\columnwidth]{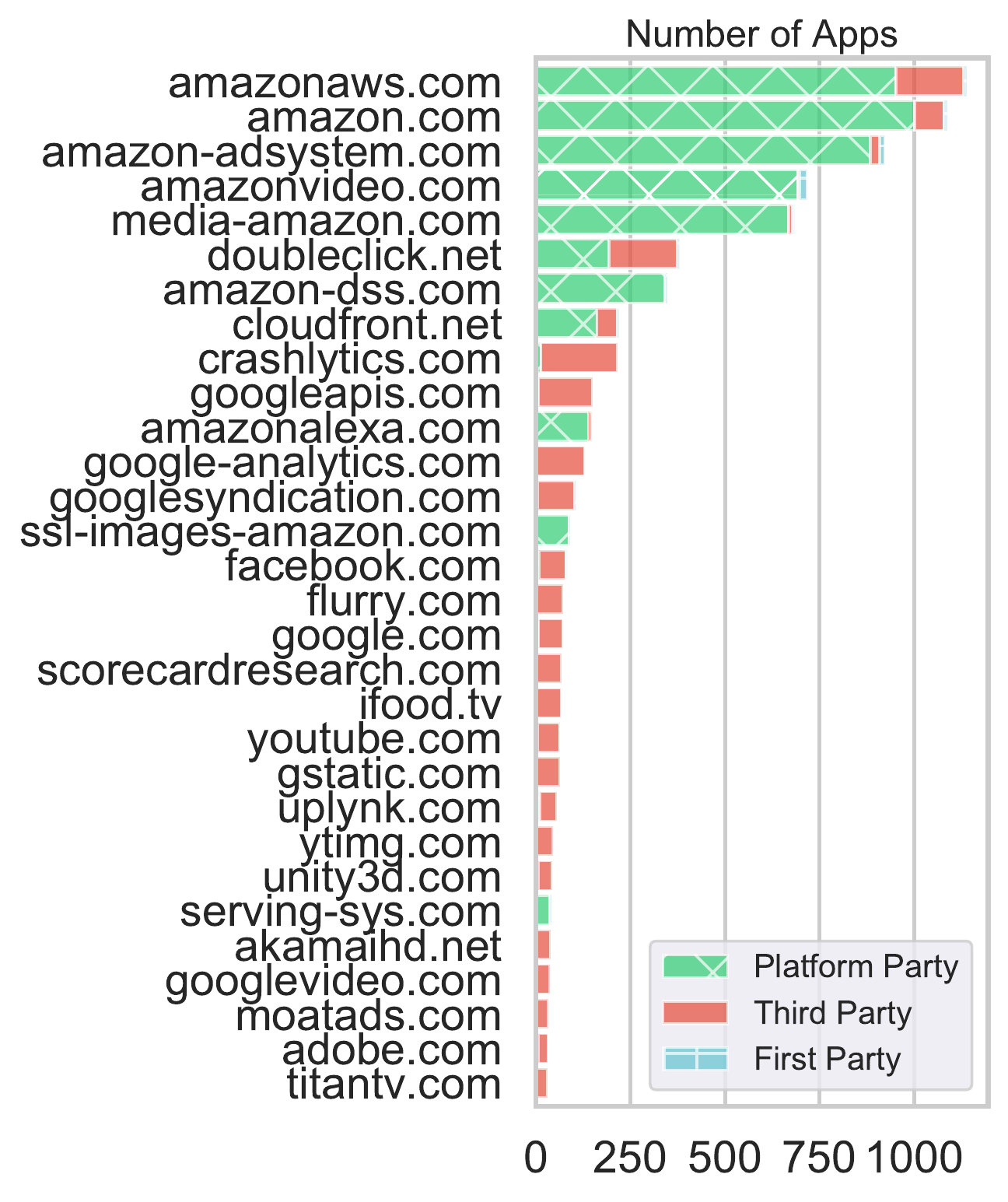}
		\caption{App penetration for the top-30 eSLDs in the \firetv{} testbed dataset.}
		\label{fig:firetv-top-hostnames-by-channels-sld}
	\end{subfigure}
	\begin{subfigure}{0.9\columnwidth}
		\centering
		\includegraphics[width=1\columnwidth]{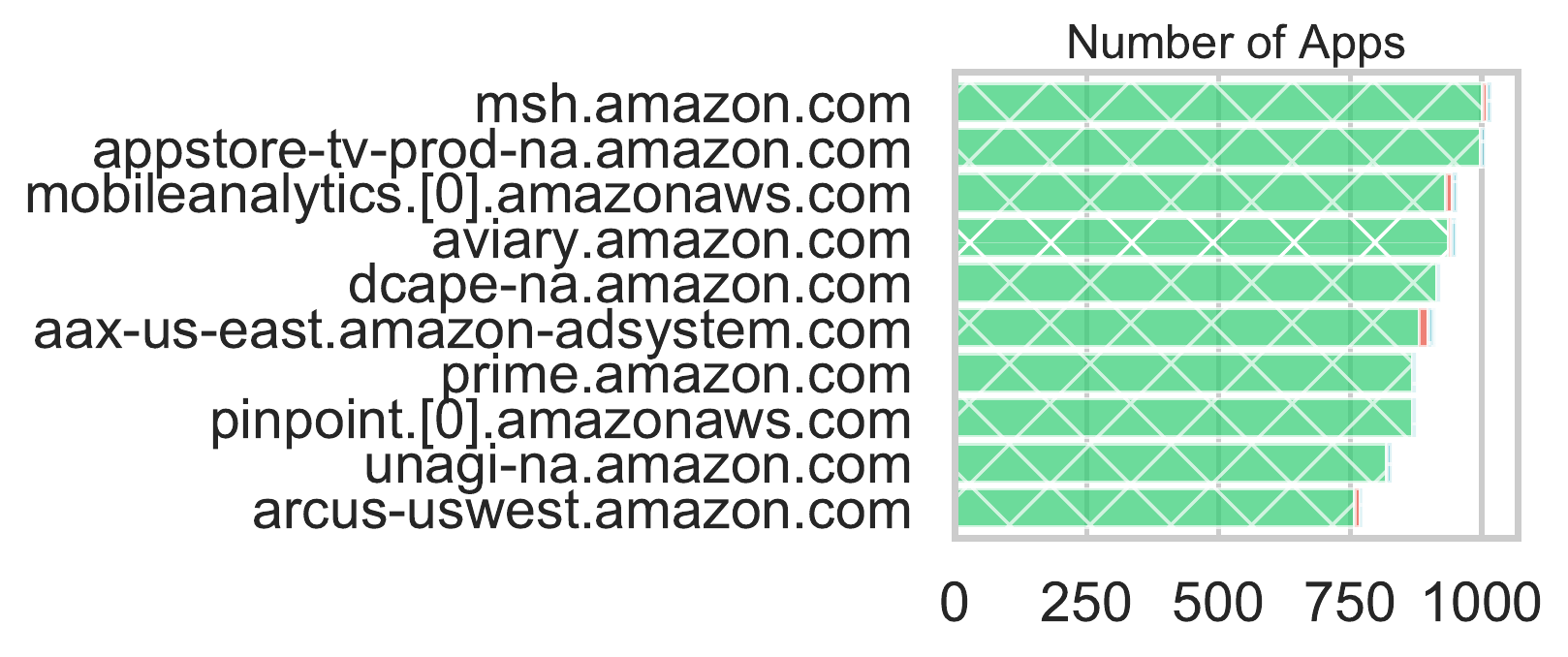}
		\caption{\firetv{} top-10 platform-specific domains.}
		\label{fig:firetv-top-hostnames-by-channels}
	\end{subfigure}
	\begin{subfigure}{0.91\columnwidth}
		\centering
		\includegraphics[width=0.96\columnwidth]{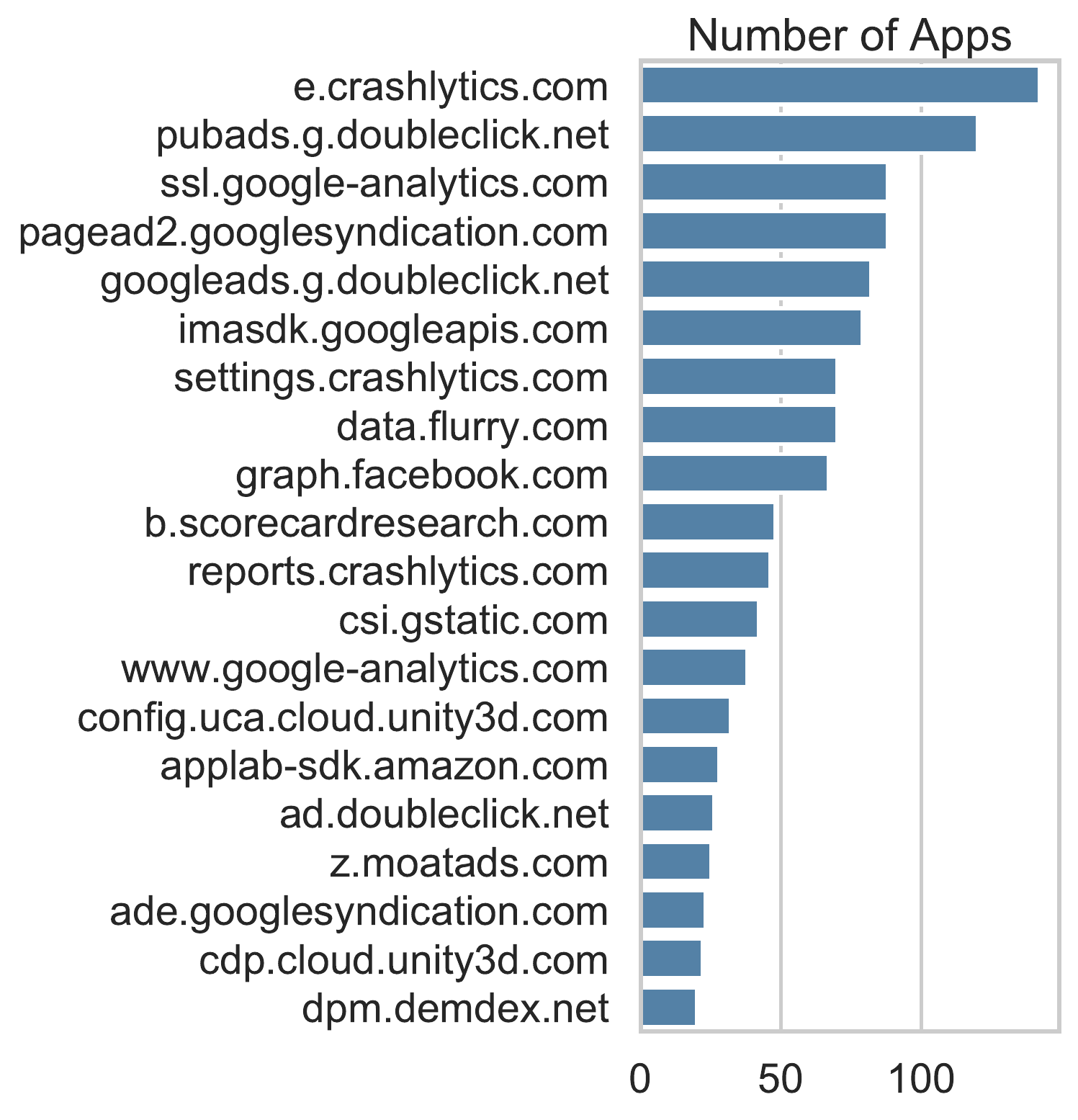}
		\caption{\firetv{} top-20 third party \ats{} domains}
		\label{fig:firetv-top-thirdparty-ats}
	\end{subfigure}
	\caption{\firetv{} testbed dataset broken down by top eSLDs, platform-specific domains, and third party \ats{} domains.}
	\label{fig:firetv-eco-all}
\end{figure}

\descr{Key Players.}
Figures~\ref{fig:roku-top-hostnames-by-channels-sld} and ~\ref{fig:firetv-top-hostnames-by-channels-sld} present the top-30 \esld{}s in terms of number of apps that contacted a subdomain of each \esld{} for each of the two platforms.	
We define an \esld{}'s \emph{app penetration} as the percentage of apps in the dataset that contact the \esld{}.
The top \esld{} for both platforms have 100\% app penetration and belong to the platform operator.
Alphabet has a strong presence in the \ats{} space of both platforms, where *.doubleclick.net, an 
ad delivery endpoint, achieves 58\% and 35\% app penetration for \roku{} and \firetv{}, respectively.
Its analytic services such as google-analytics.com and crashlytics.com also rank high on both platforms.

\descr{Platform Activities.}
Figures~\ref{fig:roku-top-hostnames-by-channels} and~\ref{fig:firetv-top-hostnames-by-channels} present the top-10 platform-specific \fqdn{}s for \roku{} and \firetv{}, respectively.
We note that both platforms use distinct advertising and tracking services.
For \roku{}, subdomains of \url{roku.com} dominate, with \url{api.sr.roku.com} and \url{cooper.logs.roku.com} being contacted by \emph{all} apps in our dataset.
\roku{} devices are known to contact log services every 30 seconds and initiate thousands of DNS lookups daily if these \fqdn{}s are blocked \cite{pihole-discourse-commonly-blacklisted, reddit-roku-2-constant-dns-requests, reddit-roku-generating-over-4000-dns-queries}.
These services are also suspected of tracking \roku{} remote key presses \cite{reddit-what-are-these-roku-domains-for}.
For \firetv{}, subdomains of \url{amazon.com}, \url{amazonaws.com}, and \url{amazon-adsystem.com} dominate.
For example, \url{mobileanalytics.us-east-1.amazonaws.com} \cite{amazon-mobile-analytics} and \url{pinpoint.us-east-1.amazonaws.com} \cite{amazon-pinpoint} analytics endpoints are contacted by 92\% and 85\% apps, respectively. 
Amazon also dominates the advertising business on its own platform, with \url{aax-us-east.amazon-adsystem.com} being contacted by 87\% apps.

\begin{table}[t!]
	\small
	\centering
	\begin{tabularx}{\columnwidth}{X|r|r|r}
		\textbf{Number of} &  \textbf{\roku{}} & \textbf{\firetv{}} &\textbf{Both} \\ \hline
		Apps exercised &  \rokutotalapps & \firetvtotalapps{} & \comappcount \\
		Fully qualified domain names (\fqdn{})  & 2191 & 1734 & 578 \\
		\fqdn{}s accessed by multiple apps & 669 & 603  & 199\\
		URL paths  & 13899 & 240713 & 74 \\
	\end{tabularx}
	\caption{Summary of the \roku{} and \firetv{} testbed datasets. The rightmost column summarizes the intersection between the two testbed datasets. For example, there are 128 apps that are present both in the Roku dataset and the Fire TV dataset.}
	\vspace{-5pt}
	\label{tab:roku-firetv-dataset}
\end{table}


\descr{Third Party Advertising and Tracking Services (\ats{}).}
We now seek to understand the \thirdparty{} \ats{} ecosystem when we strip away the platform-specific endpoints, and how they compare with the traditional mobile \ats{} ecosystem.
Figs.~\ref{fig:roku-top-thirdparty-ats} and ~\ref{fig:firetv-top-thirdparty-ats} present the top-20 \emph{\thirdparty{}} \ats{} endpoints for \roku{} and \firetv{}.
We identify \ats{} endpoints by checking if a \fqdn{} is labeled as ads or tracking by VirusTotal, McAfee, OpenDNS \cite{virustotal-homepage, mcafee-urlchecker, opendns-domaintagging}, or if it is blocked by any of the block lists considered in Section \ref{sec:eval-block-lists}.
We note that both platforms use distinct third party \ats{}. 
For example, Fig.~\ref{fig:roku-top-thirdparty-ats} shows that SpotX (spotxchange.com), which serves video ads, is a significant player in the \roku{} \ats{} space with 17\% app penetration, but only maintains 1\% app penetration for \firetv{}.
Even when considering the smaller players, we see little overlap between the two platforms, suggesting these players focus their efforts on a single platform.
For example, Kantar Group's insightexpressai.com analytics service has 7\% app penetration on the \roku{} platform, but only 0.01\% on the \firetv{} platform.

\descr{Comparing to Android \ats{} Ecosystem.}
Next, we compare the top-20 third party \ats{} endpoints in our \roku{} and \firetv{} datasets (Figs.~\ref{fig:roku-top-thirdparty-ats} and ~\ref{fig:firetv-top-thirdparty-ats}) with those reported for Android \cite{razaghpanah2018apps}.

{\em \roku{} vs. Android.}
The key third party \ats{} players in \roku{} (Fig.~\ref{fig:roku-top-thirdparty-ats}) differ substantially from the Android platform: the only overlapping \fqdn{}s are Alphabet's endpoints such as \url{tpc.googlesyndication.com} and subdomains of \url{doubleclick.net}.
While Alphabet has a strong foothold in both \roku{} and Android \ats{} ecosystems, it is less significant for \roku{}~(9 out 20 of the top \roku{} \thirdparty{} ATS FQDNs are Alphabet-owned, vs. 16 out of 20 for Android).
Furthermore, Facebook is a key player in the Android \ats{} ecosystem (graph.facebook.com is the second most popular ATS domain), but has close to zero ATS presence on the \roku{} platform; while SpotX (*.spotxchange.com) has a strong presence on \roku{} but is not among the key players for Android.
Similarly, comScore's tracking service *.scorecardresearch.com is a key player for \roku{}, but is absent for Android.
In contrast to the top \ats{} for Android, the set of top \thirdparty{} \ats{} in \roku{} is more diverse and includes smaller organizations such as Pixalate (adrta.com), Telaria (*.tremorhub.com), Barons Media (ctv.monarchards.com), and The Trade Desk (insight.adsrvr.org).

{\em \firetv{} vs. Android.}
In contrast to \roku{}, \firetv{} is much more similar to Android: we see an overlap of 9 \fqdn{}s, 7 of which are owned by Alphabet.
This is to be expected, given that \firetv{} is based off of Android, and thus natively supports the \ats{} services of Android.
For example, the Alphabet-owned analytics service Crashlytics, which is supported on Android, iOS and Unity, is widely in use on both \firetv{} and Android, but is completely absent on \roku{}. 
In contrast to \roku{}, Facebook (graph.facebook.com) and Verizon (data.flurry.com) both have a strong presence on both \firetv{} and Android.
Some of the third party \ats{} observed for \firetv{}, which were not present for Android, include comScore (*.scorecardresearch.com), Adobe (dpm.demdex.net), and Amazon (applab-sdk.amazon.com).


\begin{figure}[t!]
	\centering
	\includegraphics[width=0.80\columnwidth]{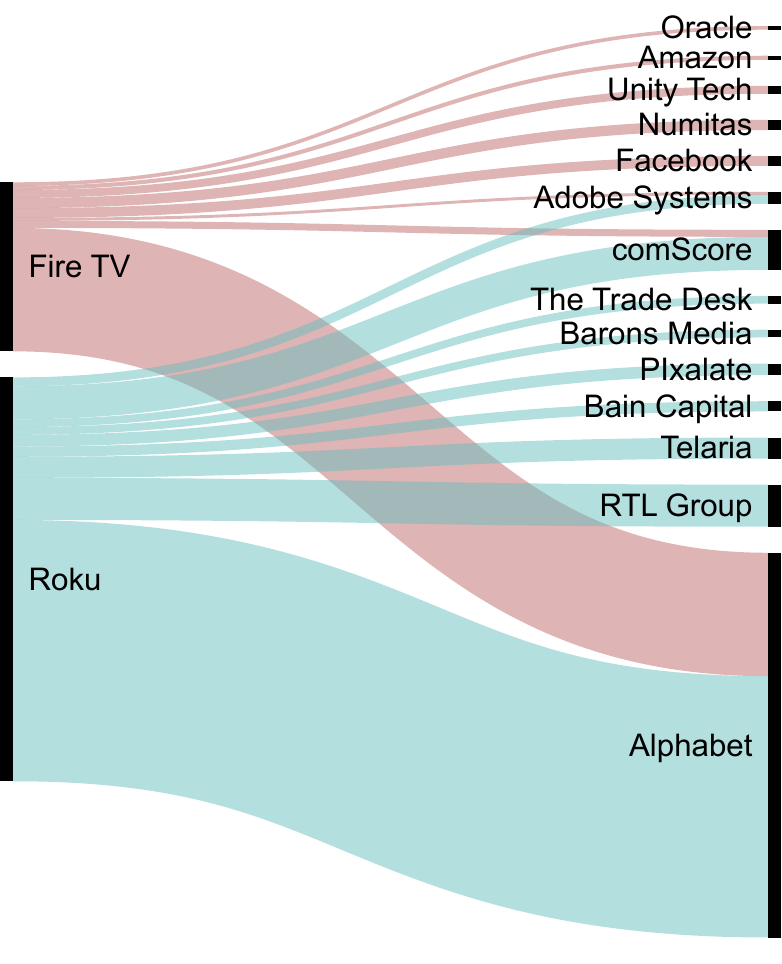}
	\caption{Mapping of \roku{} and \firetv{} devices (testbed dataset) to the parent organizations of the \fqdn{}s they contact. The width of an edge indicates the distinct number of apps that contact each organization.  The figure depicts organizations across devices that contact them most, while also highlighting smaller companies that only appear within each platform. }
	\label{fig:thirdparty-ats-testbed}
\end{figure}

\descr{Parent Organization Analysis.}
We further analyze the parent organizations for \roku{} and \firetv{} third party \ats{} endpoints in Fig. \ref{fig:thirdparty-ats-testbed} using the method described earlier in Section~\ref{sec:in-the-wild}.
Interestingly, the set of top third party organizations is rather diverse, with only a slight overlap in the shape of Adobe Systems and comScore, possibly suggesting that the remaining organizations focus their efforts on a single platform. %
\firetv{} shows Unity Tech and Facebook, suggesting that the platform has more popular gaming and social applications.  
On the other hand, \roku{} has more video content related ads being served from the Trading Desk, Telaria, and RTL Group.
Similar to in-the-wild organization analysis in Fig~\ref{fig:inthewild-flow-all}, we again note that Alphabet dominates third party \ats{} on both \roku{} and \firetv{}.

\descr{Takeaway.}
The key players of the \roku{} and \firetv{} \ats{} ecosystems differ substantially.
For example, SpotX is a relatively large player on \roku{}, but is almost absent from \firetv{}.
In contrast, Facebook has almost zero presence on \roku{}, but has a reasonable foothold on \firetv{}.
The exception is Alphabet, which has a strong \ats{} presence on both platforms.
The key \ats{} players on Android have little overlap with \roku{} but substantial overlap with \firetv{}, which is built on top of Android.
Building on our findings, we present insights for identifying platform-specific \ats{} endpoints in Section \ref{sec:eval-block-lists}.


\begin{figure*}[t!]
	\centering
	\includegraphics[width=1\linewidth]{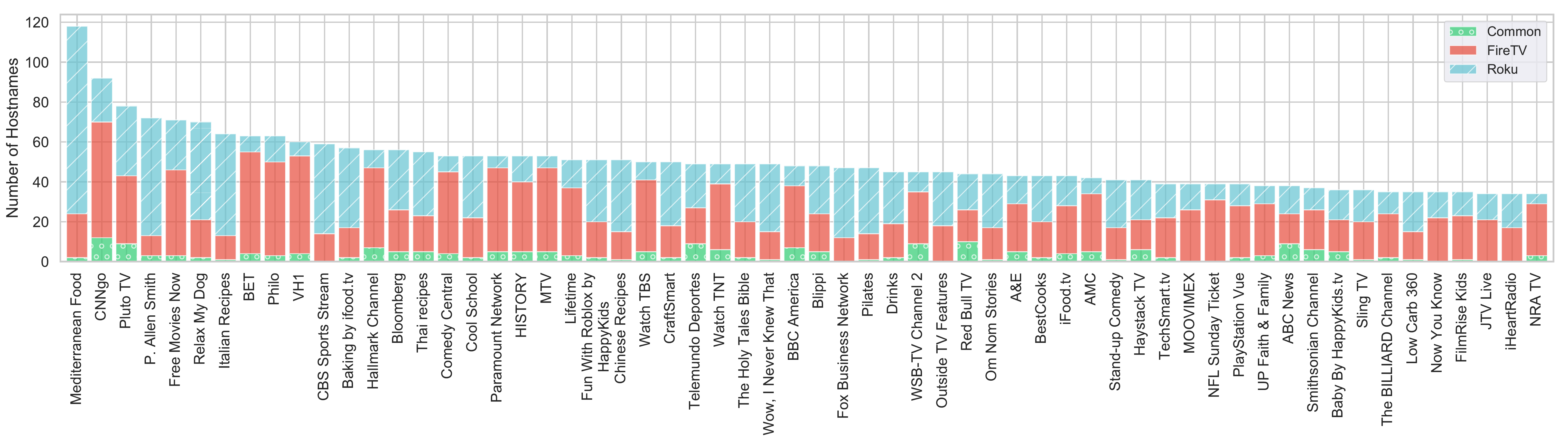}
	\caption{Top-60 common apps (apps present in both testbed datasets) ordered by the number of  hostnames that each app contacts.  Considering all common apps, there are 597 hostnames which are exclusive to \roku{} apps, 496 hostnames which are exclusive to \firetv{} apps, and 155 hostnames which are contacted by both the \roku{} and the \firetv{} version of the same app.}
	\label{fig:common-app-hostnames}
\end{figure*}

\subsection{Common Apps in \roku{} and \firetv{}} \label{sec:eco-compare-common-apps}
Next, we compare the \ats{} ecosystems of \roku{} and \firetv{} at the app-level by analyzing the traffic generated by the set of apps that appear on both platforms, referred to as common apps.
Recall from Table \ref{tab:roku-firetv-dataset} that the datasets collected using \rokutool{} and \firetvtool{} contain a total of 128 common apps.
We identified \commapps by fuzzy matching app names since they sometimes vary slightly for each platform (\eg{}, ``TechSmart.tv'' on \roku{} vs. ``TechSmart'' on \firetv{}).
We further cross-referenced with the developer's name to validate that the apps were indeed the same (\eg{}, both TechSmart apps are created by ``Future Today'').

We examine the overlapping and non-overlapping sets of endpoints contacted by the common apps.
The 128 common apps contact a total of 1248 different \fqdn{}s.
Out of these, 597 \fqdn{}s are exclusively contacted by \roku{} apps, 496 are exclusively contacted by \firetv{} apps, and only 155 \fqdn{}s are contacted by both \roku{} and \firetv{} apps.
These numbers already suggest that the two ecosystems differ substantially even for the \commapps.

Figure~\ref{fig:common-app-hostnames} reports overlapping and non-overlapping \fqdn{} for the top-60 \commapps (in terms of the number of distinct \fqdn{}s that each app contacts).
In general, the set of \fqdn{}s contacted by both the \roku{} and the \firetv{} versions of the same app is much smaller than the set of platform-specific \fqdn{}s.
%
%
From inspecting the common \fqdn{}s for the apps in Fig.~\ref{fig:common-app-hostnames}, we find that these generally include endpoints that serve content.
For example, for Mediterranean Food, the only two common hostnames are subdomains of \url{ifood.tv}, which belong to the parent organization behind the app.
This makes intuitive sense as the same app presumably offers the same content on both platforms and must therefore access the same servers to download said content.
On the other hand, the platform-specific hostnames contain obvious \ats{} endpoints such as \url{ads.yahoo.com} and \url{ads.stickyadstv.com} for the \roku{} version of the app, and \url{aax-us-east.amazon-adsystem.com} and \url{mobileanalytics.us-east-1.amazonaws.com} for the \firetv{} version of the app.
In conclusion, our analysis of \commapps that are present on both platforms reveals (to our surprise) little overlap in the \ats{} endpoints they access, which further highlights the distinct nature of the \ats{} ecosystems of the two platforms.

\section{Blocklists for Smart TVs}
\label{sec:eval-block-lists}
In this section, we are interested in detecting and mitigating \ats in the \smarttv{} ecosystem, using blocklists.  First, we evaluate four well-known DNS-based blocklists and demonstrate their ineffectiveness in blocking traffic towards \ats and preventing PII exfiltration,  across different smart TV platforms. Then, we provide insights and directions for improvement of blocklist curation specifically for \smarttv{}s.

\subsection{Evaluating Popular DNS Blocklists}

Note that unlike on-device blocking solutions that are readily available for the desktop and mobile ecosystems \cite{merzdovnik2017block}, it is generally not feasible to install ad/tracker blocking solutions directly on \smarttv{}s.
DNS-based blocking solutions such as Pi-hole \cite{pihole-homepage} are typically used to block advertising and tracking traffic from non-traditional devices in a smart home, including smart TVs \cite{pihole-discourse-what-really-happens}.
To block advertising and tracking traffic, they essentially ``blackhole'' DNS requests to known advertising and tracking domains.
Specifically, they match the domain name in a DNS request against a set of blocklists that are essentially curated hosts files that contain rules for well-known advertising and tracking services. 
We note that the well-known EasyList \cite{easylist} contains regular expressions that do fine-grained matching against URLs in HTTP requests.
In contrast, the hosts files are limited to coarse-grained matching against domain names in DNS requests. 
If the domain name is found in one of the blocklists, it is typically mapped to 0.0.0.0 or 127.0.0.1 to prevent outbound traffic to that domain \cite{pihole-docs-blocking-mode}.

\descr{Setup.} We evaluate the following blocklists:
\begin{enumerate}

\item {\bf Pi-hole Default} (\textsf{\piholelists{}}): We test blocklists included in Pi-hole's default configuration \cite{pihole-customising-sources-for-ad-lists} to imitate the experience of a typical \pihole{} user. This set has 7 hosts files including Disconnect.me ads, Disconnect.me tracking, hpHosts, CAMELEON, MalwareDomains, StevenBlack, and Zeustracker. \piholelists{} contains a total of about 133K entries.

\item {\bf The Firebog} (\textsf{\firebog{}}): We test the 9 advertising and 5 tracking blocklists recommended by  ``The Big Blocklist Collection'' \cite{firebog} to imitate the experience of an advanced \pihole{} user. This set not only includes most of the advertising and tracking lists in \pihole{}'s default configuration such as Disconnect.me ads and hpHosts, but also includes a dedicated blocklist targeting smart TVs as well as hosts versions of EasyList and EasyPrivacy. \textsf{\firebog{}} contains a total of about 162K entries.

\item {\bf Mother of all Ad-Blocking} (\textsf{\moab{}}): We test this curated hosts file \cite{moab} that targets a wide-range of unwanted services including advertising, tracking, (cookies, page counters, web bugs), and malware (phishing, spyware) to again imitate the experience of an advanced \pihole{} user. \moab{} contains a total of about 255K entries.

\item {\bf StopAd} (\textsf{\stopad{}}): We test a commercial smart TV focused blocklist by StopAd \cite{stopadtv}. This list particularly targets Android based smart TV platforms such as \firetv{}. We extract StopAd's list by analyzing its APK using Android Studio's APK Analyzer \cite{apkanalyzer}. StopAd contains a total of about 3K entries.

\end{enumerate}

We applied the aforementioned blocklists to both our in-the-wild and testbed datasets, discussed in the previous sections, and we report the results next.


\begin{table}[t!]
	\small
	\centering
	\begin{tabularx}{\columnwidth}{X r | r r r r}
		\multicolumn{2}{c|}{} & \multicolumn{4}{c}{\textbf{Block Rate (\%)}} \\
		\textbf{Platform} & \textbf{\# Domains} & \textsf{\piholelists{}} & \textsf{\firebog{}} & \textsf{\moab{}} & \textsf{\stopad{}}  \\
		\hline \multicolumn{6}{c}{\textbf{Dataset obtained ``in the wild''}} \\
		\hline
		Apple & 3179 & 10\% & 13\% & 12\% & 5\%  \\
		Samsumg & 1765 & 14\% &  19\% & 15\% & 8\% \\
		Chromecast~~~ & 1576 & 9\% & 15\% & 15\% & 5\% \\
		Roku & 2312 & 15\% & 19\% & 18\% & 7\% \\
		Vizio & 942 & 16\% &  18\% & 16\% & 11\%  \\
		LG & 627 & 45\% & 54\% & 50\% & 27\% \\
		Sony & 119 & 16\% &  24\% & 16\% & 7\% \\
		\hline \multicolumn{6}{c}{\textbf{Dataset obtained in our testbed}} \\
		\hline
		Roku & 2191 & 17\% &   22\% & 20\% & 9\% \\
		Fire TV & 1734 & 22\% &  27\% & 22\% & 9\% \\
	\end{tabularx}
	\caption{Block rates of the four blocklists in our datasets.}
	\label{tab:filter-list-performance-summary}
\end{table}

\descr{Block Rates.}
We start our analysis by comparing how much is blocked by different sets of blocklists.
Table \ref{tab:filter-list-performance-summary} compares the block rates of the aforementioned blocklists on our in the wild and testbed datasets.
Overall, we note that \textsf{\firebog{}}, closely followed by \textsf{\moab{}} and \textsf{\piholelists{}}, blocks the highest fraction of domains across all of the platforms in both in the wild and testbed datasets.
\textsf{\stopad{}} is the distant last in terms of block rate.
It is noteworthy that \textsf{\firebog{}} blocks more hostnames than \textsf{\moab{}} despite being about one-third shorter.
Comparing \textsf{\firebog{}} and \textsf{\moab{}}, we surmise that \textsf{\firebog{}} blocks more than \textsf{\moab{}} despite being smaller because \textsf{\firebog{}} includes a smart TV focused hosts file.
This finding shows that the size of a blocklist does not necessarily translate to its coverage.

\descr{Blocklist Mistakes.}
Motivated by the differences in the block rates of the four blocklists, we next compare them in terms of false negatives (FN) and false positives (FP).
   False negatives occur when a blocklist does not block requests to \ats and may result in (visually observable) ads or (visually unobservable) \pii{} exfiltration.
    False positives occur when a blocklist blocks requests that enable app functionality and may result in (visually observable) app brekage.

We first systematically quantify visually observable false positives and false negatives of blocklists by interacting with a sample of apps from our testbed datasets and manually coding for presence of ads and app breakage.
We sample 10 \roku{} apps and 10 \firetv{} apps, including the top-4 free apps, three apps that are present on both platforms, and an additional three randomly selected apps. 
We test each app five times: one time without any blocklist and four times where we individually deploy each of the aforementioned blocklists.
During each experiment, we attempt to trigger ads by playing multiple videos and/or live TV channels and fast-forwarding through video content, and we take note of any visually observable functionality breakage (due to false positives) and missed \ats (due to false negatives).
We differentiate between minor and major functionality breakage as follows: \textit{minor breakage} when the app's main content remains available but the application suffers from minor user interface glitches or occasional freezes; and \textit{major breakage} when the app's content becomes completely unavailable or the app fails to launch.

\begin{table*}[t!]
	\small
	\begin{tabularx}{\textwidth}{ l | l | X | c  c | c  c | c  c | c  c | c  c }
		\multicolumn{2}{c}{} & \multicolumn{1}{c|}{\textbf{App Name}} & \multicolumn{2}{c|}{{No List}} & \multicolumn{2}{c|}{\textsf{\piholelists{}}} & \multicolumn{2}{c|}{\textsf{\firebog{}}} & \multicolumn{2}{c|}{\textsf{\moab{}}} & \multicolumn{2}{c}{\textsf{\stopad{}}} \\
		\multicolumn{2}{c}{} & & \multicolumn{1}{P{0.5cm}}{No Ads} & \multicolumn{1}{P{1.16cm}|}{No Breakage} & \multicolumn{1}{P{0.5cm}}{No Ads} & \multicolumn{1}{P{1.16cm}|}{No Breakage} & \multicolumn{1}{P{0.5cm}}{No Ads} & \multicolumn{1}{P{1.16cm}|}{No Breakage} & \multicolumn{1}{P{0.5cm}}{No Ads} & \multicolumn{1}{P{1.16cm}|}{No Breakage} & \multicolumn{1}{P{0.5cm}}{No Ads} & \multicolumn{1}{P{1.16cm}}{No Breakage} \\
		\hline
		\parbox[t]{1mm}{\multirow{10}{*}{\rotatebox[origin=c]{90}{\textbf{\roku{}}}}} & \parbox[t]{1mm}{\multirow{3}{*}{\rotatebox[origin=c]{90}{Common}}} & \scriptsize{Pluto TV} & \xmark{} & \scalecheck{} & \xmark{} & \scalecheck{} & \xmark{} & \scalecheck{} & \xmark{} & \scalecheck{} & \xmark{} & \scalecheck{} \\
		& & \scriptsize{iFood.tv} & \xmark{} & \scalecheck{} & \scalecheck{} & \scalecheck{} & \scalecheck{} & \scalecheck{}	& \scalecheck{} & \scalecheck{} & \xmark{} & \scalecheck{}  \\
		& & \scriptsize{Tubi} & \scalecheck{} & \scalecheck{} & \scalecheck{} & \scalecheck{} & ---  &  {\xmarkbold{}}	& \scalecheck{} & \scalecheck{} & \scalecheck{} & \scalecheck{} \\
		\cline{2-13}
		& \parbox[t]{1mm}{\multirow{4}{*}{\rotatebox[origin=c]{90}{Top}}} & \scriptsize{YouTube} & \xmark{} & \scalecheck{} & \xmark{} & \scalecheck{} & \xmark{} & \scalecheck{}	& \xmark{} & \scalecheck{} & \xmark{} & \scalecheck{} \\
		& & \scriptsize{CBS News Live} & \xmark{} & \scalecheck{} & \scalecheck{} &  {\textbf{\xmarkbold{}}} & \scalecheck{}  &  {\xmarkbold{}} & \scalecheck{} &  {\xmarkbold{}} & \scalecheck{} & \scalecheck{} \\
		& & \scriptsize{The Roku Channel} & \xmark{}	& \scalecheck{} & \scalecheck{} & \scalecheck{} & \scalecheck{} &  {\xmarkbold{}} & \scalecheck{} & \scalecheck{} & \xmark{} & \scalecheck{} \\
		& & \scriptsize{Sony Crackle} & \xmark{}	& \scalecheck{} & \xmark{} & \scalecheck{} & \scalecheck{}  & \scalecheck{} & \xmark{} & \scalecheck{} & \xmark{} & \scalecheck{}  \\
		\cline{2-13}
		& \parbox[t]{1mm}{\multirow{3}{*}{\rotatebox[origin=c]{90}{Random}}} & \scriptsize{WatchFreeComedyFlix} & \xmark{} & \scalecheck{} & \scalecheck{} & \scalecheck{} & \scalecheck{} & \scalecheck{} & \xmark{} & \scalecheck{} & \xmark{} & \scalecheck{} \\
		& & \scriptsize{Live Past 100 Well} & \scalecheck{}	& \scalecheck{} & \scalecheck{} & \scalecheck{} & \scalecheck{} & \scalecheck{} & \scalecheck{} & \scalecheck{} & \scalecheck{} & \scalecheck{}  \\
		& & \scriptsize{SmartWoman} & \xmark{} & \scalecheck{} & \scalecheck{} & \scalecheck{} & \scalecheck{} & \scalecheck{} & \scalecheck{} & \scalecheck{} & \scalecheck{} & \scalecheck{}   \\
		\hline
		\parbox[t]{1mm}{\multirow{10}{*}{\rotatebox[origin=c]{90}{\textbf{\firetv{}}}}} & \parbox[t]{1mm}{\multirow{3}{*}{\rotatebox[origin=c]{90}{Common}}} &  \scriptsize{Pluto TV} & \xmark{} & \scalecheck{} & \xmark{} & \scalecheck{} & \xmark{} & \scalecheck{} & \xmark{} &  {\xmark{}} & \xmark{} &  {\xmarkbold{}}  \\
		& & \scriptsize{iFood.tv} & \xmark{} & \scalecheck{} & \scalecheck{} & \scalecheck{} & --- &  {\xmarkbold{}}	& --- &  {\xmarkbold{}} & \scalecheck{} & \scalecheck{}  \\
		& & \scriptsize{Tubi} & \scalecheck{} & \scalecheck{} & \scalecheck{} & \scalecheck{} & \scalecheck{} & \scalecheck{} & \scalecheck{} & \scalecheck{} & \scalecheck{} & \scalecheck{} \\
		\cline{2-13}
		& \parbox[t]{1mm}{\multirow{4}{*}{\rotatebox[origin=c]{90}{Top}}} & \scriptsize{Downloader} & \scalecheck{} & \scalecheck{} & \scalecheck{} & \scalecheck{} & \scalecheck{} & \scalecheck{} & \scalecheck{} & \scalecheck{} & \scalecheck{} & \scalecheck{} \\
		& & \scriptsize{The CW for Fire TV} & \xmark{} & \scalecheck{} & --- & {\xmarkbold{}}  & --- &  {\xmarkbold{}}	& --- &  {\xmarkbold{}} & \xmark{} & \scalecheck{} \\
		& & \scriptsize{FoxNow} & \xmark{} &  \scalecheck{} & --- &  {\xmarkbold{}} & --- &  {\xmarkbold{}} & \xmark{} & \scalecheck{} & \xmark{} & \scalecheck{} \\
		& & \scriptsize{Watch TNT} & \scalecheck{} & \scalecheck{} & \scalecheck{} & \scalecheck{} & \scalecheck{} & \scalecheck{}	& \scalecheck{} & \scalecheck{} & \scalecheck{} & \scalecheck{} \\
		\cline{2-13}
		& \parbox[t]{1mm}{\multirow{3}{*}{\rotatebox[origin=c]{90}{Random}}} & \scriptsize{KCRA3 Sacramento} & \xmark{} &  \scalecheck{} & \scalecheck{} & \scalecheck{} & --- &  {\xmarkbold{}}	 & --- &  {\xmarkbold{}} & \xmark{} &  {\xmark{}} \\
		& & \scriptsize{Watch the Weather Channel} & \xmark{} & \scalecheck{} & \scalecheck{} & \scalecheck{} & \scalecheck{} & \scalecheck{} & \xmark{} & \scalecheck{} & \scalecheck{} & \scalecheck{} \\
		& & \scriptsize{Jackpot Pokers by PokerStars} & \scalecheck{} & \scalecheck{} & \scalecheck{} & \scalecheck{} & \scalecheck{} & \scalecheck{}	 & \scalecheck{} & \scalecheck{} & \scalecheck{} &  {\xmark{}} \\
	\end{tabularx}
	\caption{Missed \ats and functionality breakage for different blocklists when employed during manual interaction with 10 \roku{} apps and 10 \firetv{} apps. For ``No Ads'', a checkmark (\protect\scalecheck{}) indicates that no ads were shown during the experiment, a cross (\protect\xmark{}) indicates that some ad(s) appeared during the experiment, and a dash (---) indicates that breakage prevented interaction with the app altogether.
	For ``No Breakage'', a checkmark (\protect\scalecheck{}) indicates that the app functioned correctly, a cross (\protect\xmark{}) indicates minor breakage, and a bold cross (\protect\xmarkbold{}) indicates major breakage.
	}
	\label{tab:false-positives}
\end{table*}

\descr{Functionality Breakage vs. Missed \ats.}
Table \ref{tab:false-positives} summarizes the results of our manual analysis for functionality breakage and missed ads.
No missed ads is indicated using a check mark and missed ads is indicated using a cross.
No breakage is indicated using a check mark, minor breakage is indicated using a cross, and major breakage is indicated using a bolded cross.
Overall, we find that none of the blocklists are able to block ads from all of the sampled apps while avoiding breakage.
In particular, none of the blocklists are able to block ads in YouTube and Pluto TV (available on both \roku{} and \firetv{}).
Across different lists, \textsf{\piholelists{}} seems to achieve the best balance between blocking ads and preserving functionality.

For Roku, \textsf{\piholelists{}} and \textsf{\firebog{}} perform similarly.
While \textsf{\firebog{}} is the only list that blocks ads in Sony Crackle, both lists miss ads in YouTube and Pluto TV.
\textsf{\firebog{}} majorly breaks three apps while \textsf{\piholelists{}} only majorly breaks one app.
\textsf{\moab{}} is unable to block ads in four apps and majorly breaks only one app.
\textsf{\stopad{}} does not cause any breakage but is unable to block ads in six apps.

For Fire TV, \textsf{\piholelists{}} again seems to be the most effective at blocking ads while avoiding breakage, but is still unable to block ads in one app (Pluto TV) and majorly breaks two apps.
\textsf{\firebog{}} is also unable to block ads in Pluto TV, but majorly breaks four apps.
\textsf{\moab{}} is unable to block ads in three apps and majorly breaks three apps (one minor).
\textsf{\stopad{}} is unable to block ads in four apps and majorly breaks one app (two minor).


\descr{Takeaway.}
Unfortunately, all blocklists suffer from a non-trivial amount of visually observable false positives and false negatives.
Some blocklists (\eg{}, \textsf{PD} and \textsf{TF}) are clearly more effective than others. Interestingly, \textsf{SATV}, which is curated specifically for \smarttv{}s,  did not perform well. 

%
%

\subsection{Beyond Domains}
In this section, we look at \ats characteristics beyond just the destination domains.
In particular, we observe that the more apps that contact a single destination, the more likely it is for that destination to be an ATS. 
This is intuitive and consistent with a similar observation previously made in the mobile ecosystem \cite{razaghpanah2018apps}.
This observation may also aid blocklist curators in identifying candidate block rules.

\begin{table}[t!]
	\small
	\begin{tabular}{l c c c c }
		\textbf{Hostname}  & \textsf{\piholelists{}} & \textsf{\firebog{}} & \textsf{\moab{}} & \textsf{\stopad{}} \\
		\hline
		p.ads.roku.com 										  &  \xmark{}  &   \xmark{} & \xmark{} & \xmark{} \\
		ads.aimitv.com 										   &  \xmark{} &   \xmark{} & \xmark{} & \xmark{} \\
		adtag.primetime.adobe.com 							  &  \xmark{} &   \xmark{} & \xmark{} & \xmark{}  \\
		ads.adrise.tv 					    & \xmark{} &   \scalecheck & \xmark{} & \xmark{} \\

		ads.samba.tv		 	& \xmark{} &   \scalecheck & \xmark{} & \xmark{}   \\
		tracking.sctv1.monarchads.com		  	& \xmark{} &   \scalecheck & \xmark{} & \xmark{}   \\
		data.ad-score.com	 	&  \scalecheck &   \scalecheck & \xmark{} & \xmark{}   \\
	\end{tabular}
	\caption{Examples of false negatives for the four DNS-based blocklists found using keywords searches.}
	\label{tab:false_negatives}
\end{table}

We first use simple keywords such as ``ad'', ``ads'' and ``tracking'' to shortlist obvious \ats{} domains in our datasets. 
While keyword search is not perfect, this simple approach identified several obvious false negatives, a few of which are shown in Table \ref{tab:false_negatives}.
For example, \url{p.ads.roku.com} and \url{adtag.primetime.adobe.com} are advertising/tracking related domains which are not blocked by any of the lists.
A few domains such as \url{ads.samba.tv} are only blocked by the \textsf{\firebog{}} block list.
Finally, \url{data.ad-score.com} is blocked by \textsf{\piholelists{}} and \textsf{\firebog{}}, but not \textsf{\moab{}} and \textsf{\stopad{}}.

We observe that many of these false negatives (i.e., missed \ats  domains) are contacted by multiple apps in our testbed datasets.
For example, \url{p.ads.roku.com} is accessed by more than 100 apps in our Roku testbed dataset.
To gain further insight into potential false negatives, we study whether the likelihood of being blocked is impacted by the number of apps that access a particular domain.
%
Figure \ref{fig:blocking-as-func-of-hostname-popularity} plots the block rates for the union of the four blocklists as a function of \fqdn{}s' occurrences across apps in our testbed datasets.
We note that the block rate of a domain substantially increases as it starts to appear across multiple apps.
For example, a domain's block rate almost doubles when it is contacted by multiple apps.
Domains that are contacted by multiple different apps are therefore more likely to belong to \thirdparty{} \ats{} libraries included by smart TV apps.

\begin{table*}[t!]
	\small
	\centering
	\begin{tabularx}{\textwidth}{ l | l l l l | l l l l}
		\multicolumn{1}{c|}{\textbf{PII}} & \multicolumn{4}{c|}{\textbf{Roku Testbed Dataset}} & \multicolumn{4}{c}{\textbf{Fire TV Testbed Dataset}}\\
		& First Party  & Third Party  & Platform &Total  &
		First Party & Third Party  & Platform & Total  \\
		\hline
		Advertising ID & 1899 (91\%)  & 2896 (98\%) & 0 & 4853 (94\%)  & 408 (21\%) & 1507 (81\%) & 7648 (8\%) & 9575 (20\% )  \\
		Serial Number & 743 (8\%) & 1095 (73\%) & 31 (0\%) & 1897 (46\%) & 157 (0\%) & 2864 (2.6\%) & 10411 (0\%) & 13450 (0.6\%)  \\
		Device ID & 0 & 0 & 0 &  0 & 728 ( 0\%) & 2311 (35\%) & 8578 (0\%) & 13302 (7\%) \\
		Account Name & 6 (33\%) & 0 & 0 & 8 (25\%) & 2 (0\%) & 52 (100\%) & 1 (0\%) & 55 (95\%)  \\
		MAC Address & 0 & 0 & 0 & 0 & 0 & 38 (100\%) & 0 & 38 (100\%)  \\
		Location& 55 (100\%) & 40 (100\%) & 0 & 95 (100\%)  & 0 & 225 (99\%) & 12 (100\%) & 237 (99\%)  \\
	\end{tabularx}
	\caption{{\bf Number of Exposures (\% Blocked):}  Number of times that a PII was exposed and the percentage of times that the exposure was successfully blocked by at least one blocklist. We report the total number of PII exposures found in our testbed datasets, and we further distinguish w.r.t. the packet's destination  being  first  party,  third party  or platform related, as defined in Sec.~\ref{sec:eco-compare}.}
	\label{tab:pii-firetv-roku-by-party}
\end{table*}

\subsection{\pii{} Exposures}
\label{sec:pii-exposure}

In this section, we consider exposure of personally identifiable information (\pii) and we evaluate the effectiveness of blocklists in preventing that.   
We define ``\pii exposure'' as the transmission of any PII from the smart TV device to any Internet destination.
We identify PIIs (such as advertising ID and serial number) through each platform's settings menus and packaging.
Since trackers are known to encode or hash PIIs \cite{englehardt2018never}, we compute the MD5 and SHA1 hashes for each of the PII values.
We then search for  these PIIs in the HTTP headers and URI. %

\begin{figure}[t!]
	\centering
	\includegraphics[width=\columnwidth]{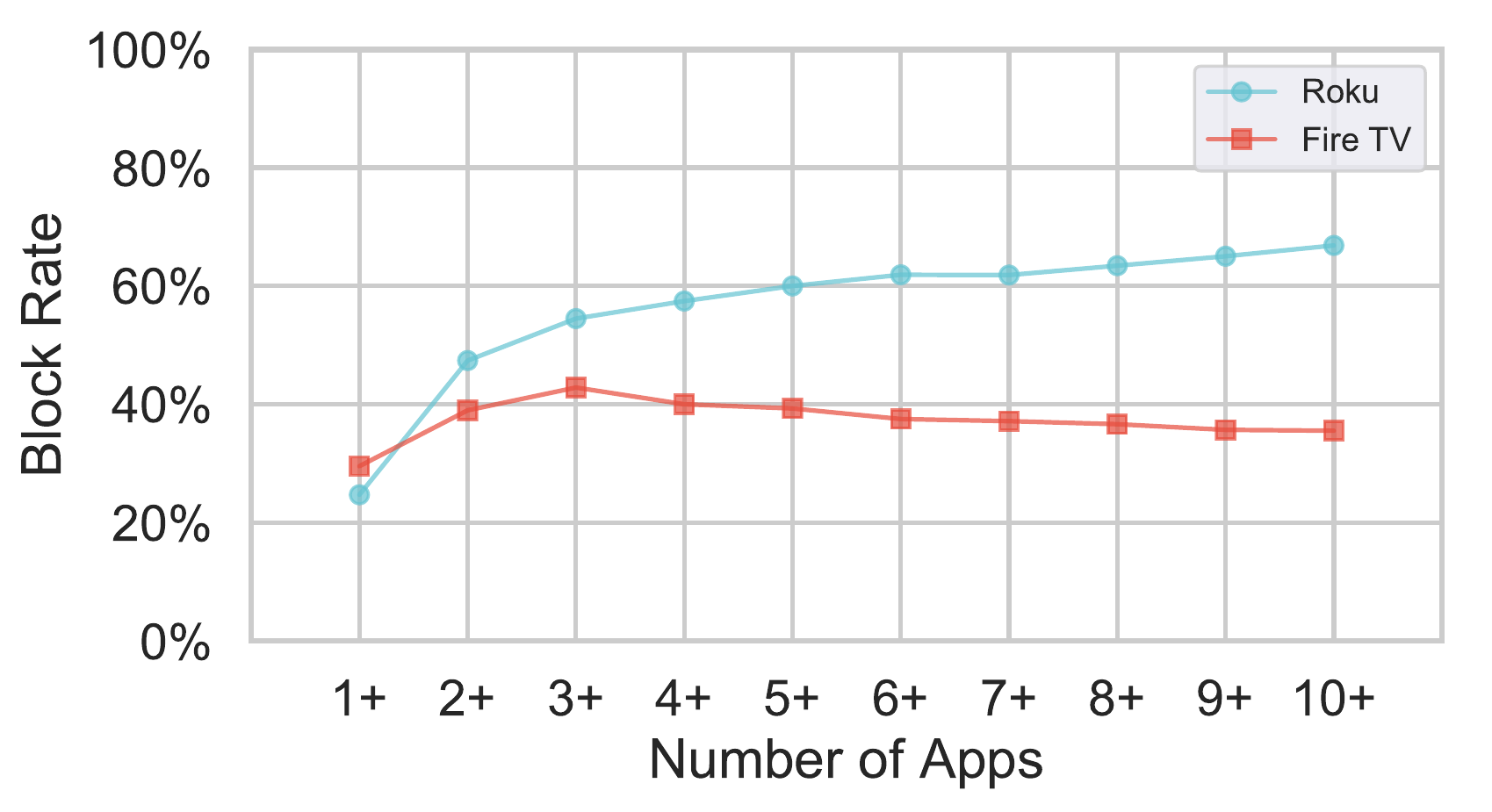}
	\caption{Block rates as a function of the number apps that contact an \fqdn{}. For the horizontal axis, ``2+'' represents the set of \fqdn{}s that are contacted by 2 or more apps.  For Roku, the more apps that contact an \fqdn{}, the more likely it is that that \fqdn{} is an \ats{}, according to the blocklists. The same is not true for \firetv{} because platform services start to dominate the set of \fqdn{}s, and platform services are often not blocked.}
	\label{fig:blocking-as-func-of-hostname-popularity}
\end{figure}

%
%

Some \pii exposures may be legitimate, in the sense that  the user may  have given permission to the app to access the PII and the network; or that the \pii sent by the app is needed for the functionality of the app. For example, Roku apps may use \pii{} to enable personalized content in the absence of logins.  However, \thirdparty{} (\ats) libraries automatically inherit app permissions and may use \pii for tracking unrelated to app functionality.  For example, in Fire TV, we observed examples where static \pii{}s (serial number) are sent alongside dynamic ones (advertising ID), which allows re-linking of users to dynamic \pii{}s (even after they have been reset).

As a first step for distinguishing between ``good'' or ''bad'' PII transmission (which is a problem on its own and out of scope for this paper), we adopt a simple approach similar to~\cite{merzdovnik2017block} that treats data exfiltration to third parties as a higher threat to privacy. We further distinguish between \pii sent to \firstparty{}, \thirdparty{}, and platform services, as defined in Sec.~\ref{sec:eco-compare}. 
	\pii{} sent to first parties are generally warranted as they have functional purposes such as personalization of content (\eg, keep track of where the user paused a video or serve content specific to the user's region).
	On the other hand, \pii{} sent to third parties do not typically have a functional purpose.
	This extends to cases where the app retrieves its content through a third party CDN as the personalization could be achieved by first sending the \pii{} to the first party server which could then respond to the app with the CDN URL for the content to be retrieved.

Table \ref{tab:pii-firetv-roku-by-party} reports the number of \pii{} exposures found in our \roku{} and \firetv{} testbed datasets.
Recall from Section \ref{sec:popular-app-testing} that we can analyze HTTP information even for encrypted flows in \firetv{}, but can only analyze unencrypted flows in Roku. Thus, we are able to observe more PII exposures for \firetv{} than \roku{}, which results in higher numbers on the right than on the left side of Table~\ref{tab:pii-firetv-roku-by-party}.

	For {\bf \roku{}}, we observe that the device's serial number is often exposed to first and third parties.
	When sent to first parties, this identifier can enable personalization  as many \roku{} apps do not require login.
	The blocklists seem to capture this functional purpose well: the block rates  for serial number are very low (8\%), when it is sent to first parties, and much higher (73\%) when it is sent to third parties. 
	There are very few exposures of serial number to the platform, none of which is captured by the blocklists. 
	Upon closer inspection, we found that  these platform exposures originate from the flickr app (developed by \roku{}) and go to \url{rokucom-link.appspot.com}. They seem to link the \roku{} device with the flickr service for analytics purposes as the URI path ``/getRegResult?...'' encodes the \pii{} alongside \roku{} as a partner and flickr as the service.

	In contrast, for {\bf \firetv{}},  blocklists are unable to block third party exposures of more prevalent \pii{}s such as serial number (only 2.6\% blocked) and device ID (35\%) while performing well for advertising ID (81\%).	
	While exposure of serial number or device ID may be utilized for software updates, we discover that they are multipurpose, as the vast majority of these exposures to Amazon-owned endpoints go to \url{aviary.amazon.com} with a URI path of ``/GetAds''.
	Furthermore, these two hardwired identifiers are often sent \emph{alongside} the advertising ID, which allows third parties to associate an old advertising ID (users can reset their advertising ID) to the new value by joining on serial number and device ID. 

	{\em Leveraging Missed \pii{} Exposures to Improve Blocklists.} 
	The unblocked \pii{} exposures indicate another direction for improving blocklist curation for \smarttv{}s. By deploying tools such as \rokutool{} and \firetvtool{} and searching the network traces for \pii{}s, blocklist curators can generate candidate rules that can then be examined manually.
	Using this approach, we identified 38 domains in the \roku{} dataset and 30 in the \firetv{} dataset that receive \pii{}, but were not blocked by any list. These numbers are conservative as we exclude location and account name that are likely to be used for legitimate purposes, such as logging in or serving location-based content.
	These domains include obvious \ats{}  such as \url{ads.aimitv.com} and \url{ads.ewscloud.com}. 
	Another noteworthy mention is \url{hotlist.samba.tv}:
	Samba TV uses Automatic Content Recognition to provide content suggestions on \smarttv{}s, but this comes at the cost of targeted advertising that even propagates onto other devices in the home network \cite{sambatv-nyt}.


\section{Conclusion \& Directions  \label{sec:conclusion}}

In this paper, we performed a comprehensive measurement study of the smart TV advertising and tracking services (\ats{}) ecosystem. While \smarttv{} platforms are becoming increasingly popular, their \ats{} ecosystems have not yet received as much attention from the privacy community as the \ats{} ecosystems of the mobile platform and the web.
To that end, we analyzed and compared: (i) a realistic but small  {\em in-the-wild} dataset  (57 \smarttv{} devices of 7 different platforms, with coarse flow-level information) and (ii)  two large {\em testbed} datasets  (top-1000 apps on \roku{} and \firetv{}, tested systematically, with granular per app and  packet-level information).

We showed that smart TVs generate a substantial amount of traffic towards \ats.  Our results also revealed that advertising and tracking is platform-specific; even common apps across different platforms tend to contact more platform-specific than common domains. Our evaluation of  four sets of state-of-the-art DNS blocklists for smart TVs showed varying block rates, with all lists suffering from both false positives (resulting in functionality breakage) and false negatives (resulting in missed ads and trackers). Even the commercial StopAd designed for smart TVs did not work well. We demonstrated that these blocklists are less effective in blocking PII sent to platform endpoints than to third parties, both of which perform tracking in smart TVs. Finally, we offered two insights to help blocklist curators identify false negatives, \ie{}, domains that are contacted by multiple apps or domains that collect \pii{}.

In summary, our work establishes that (i) the smart TV \ats{} ecosystem is fragmented across different smart TV platforms; (ii) the smart TV \ats{} ecosystem is different from the mobile \ats ecosystem; and (iii) DNS-based blocklists are currently not able to effectively filter \ats{} for \smarttv{}s. These findings motivate  more research to  further understand \smarttv{}s and to develop  privacy-enhancing solutions specifically designed for each smart TV platform. For example, more research is needed to curate accurate, fine-grained (compared to DNS-based), and platform-specific blocklists. 
 To foster further research along this direction, we plan to make our tools, \rokutool~ and \firetvtool, and datasets from testing the top-1000 \roku{} and \firetv{} apps publicly available.

\section*{Acknowledgements}
This work is supported in part by the National Science Foundation under grant numbers 1815666 and 1815131, Seed Funding by UCI VCR, and Minim. The authors would like to thank M. Hammad Mazhar and the team at Minim for their help with collecting and analyzing smart TV traffic in the wild.

\bibliographystyle{unsrt}
\bibliography{refs}

\appendix
\section{In-the-Wild Dataset}
\label{sec:inthewild-complete-appendix}



\begin{figure}[h!]
	\begin{subfigure}{0.85\columnwidth}
		\centering
		\includegraphics[width=0.85\linewidth]{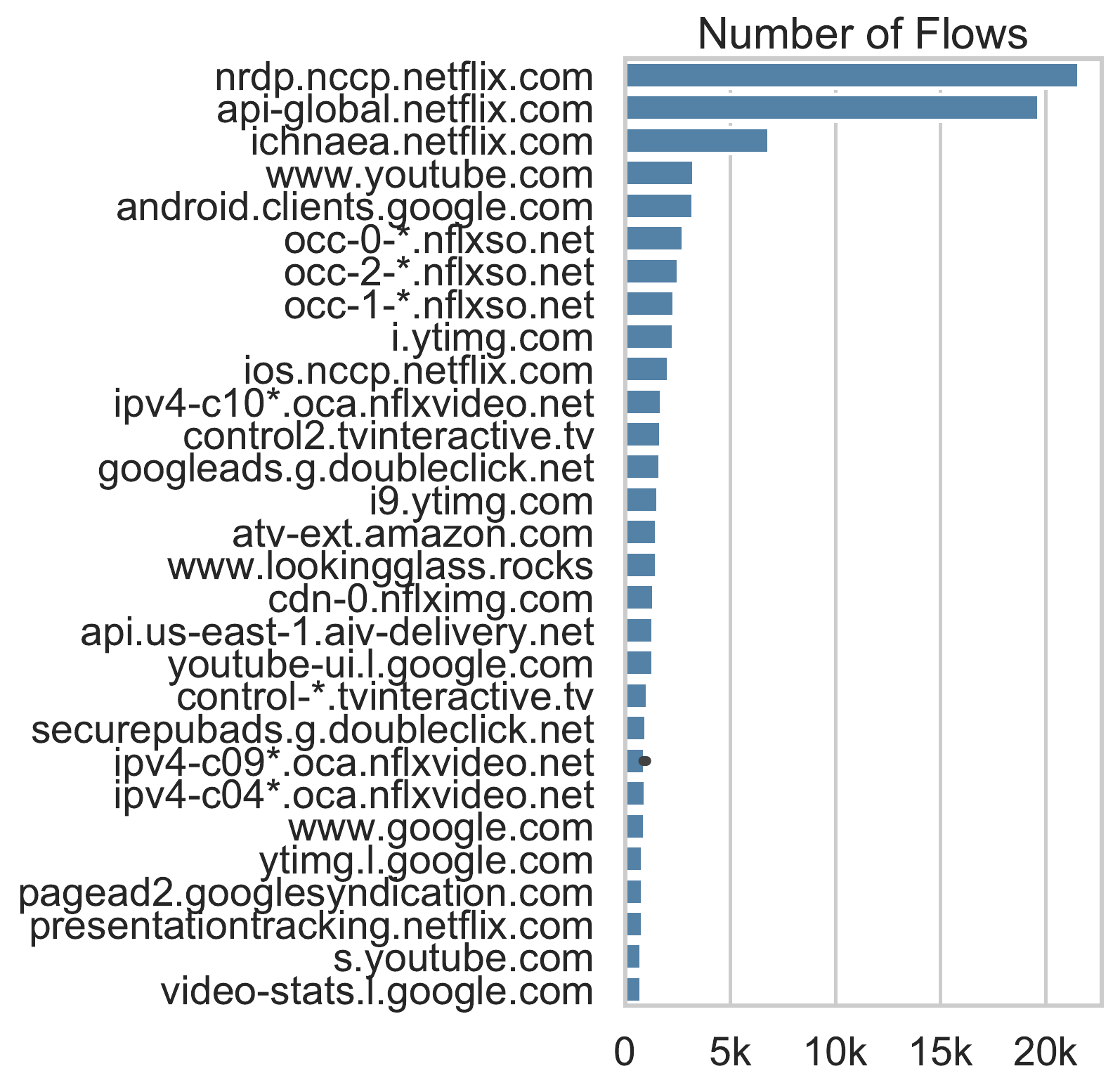}
		\caption{Vizio}
	\end{subfigure}
	\begin{subfigure}{0.85\columnwidth}
		\centering
		\includegraphics[width=0.85\linewidth]{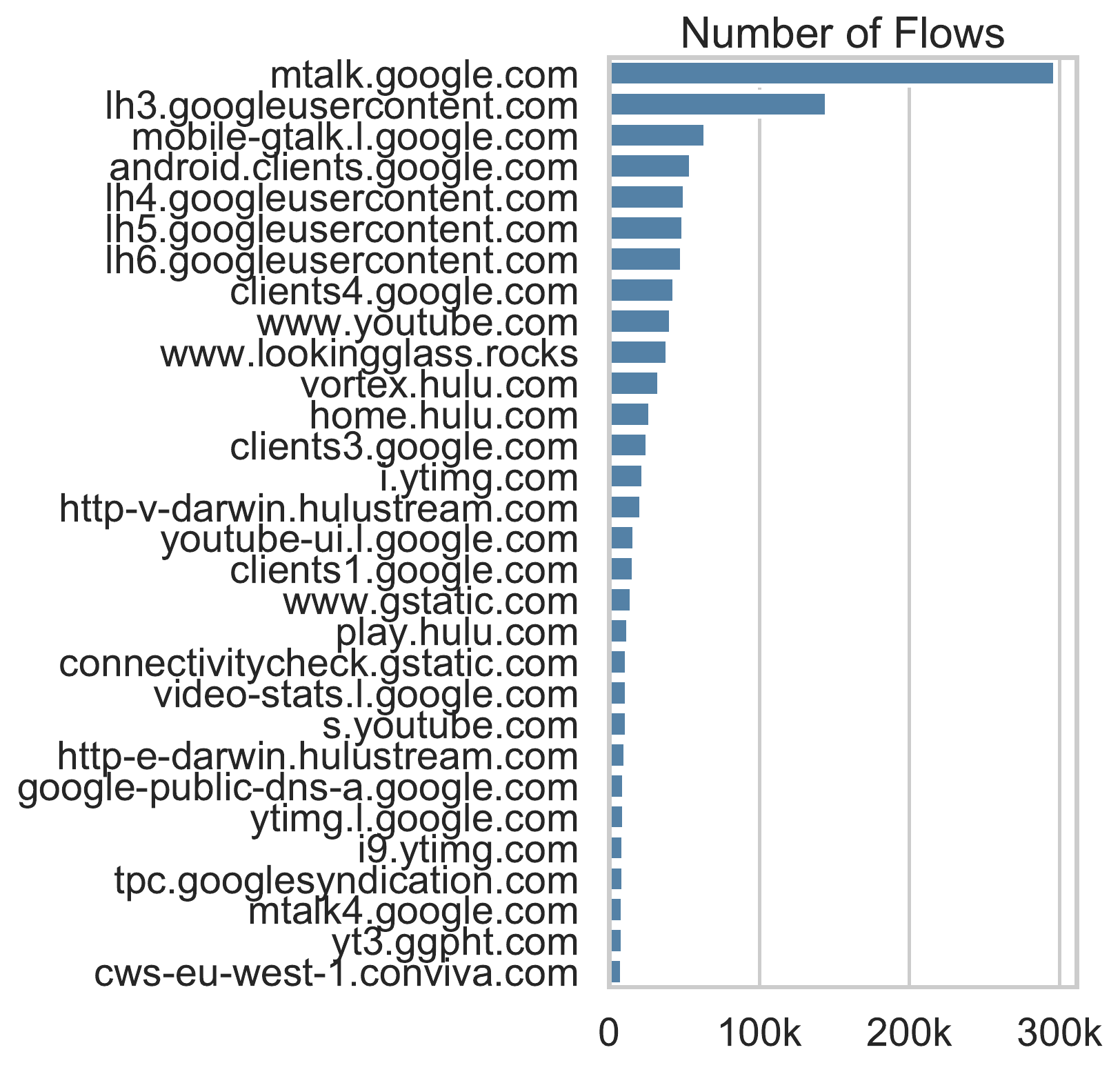}
		\caption{Chromecast}
	\end{subfigure}


	\begin{subfigure}{0.80\columnwidth}
		\centering
		\includegraphics[width=0.85\linewidth]{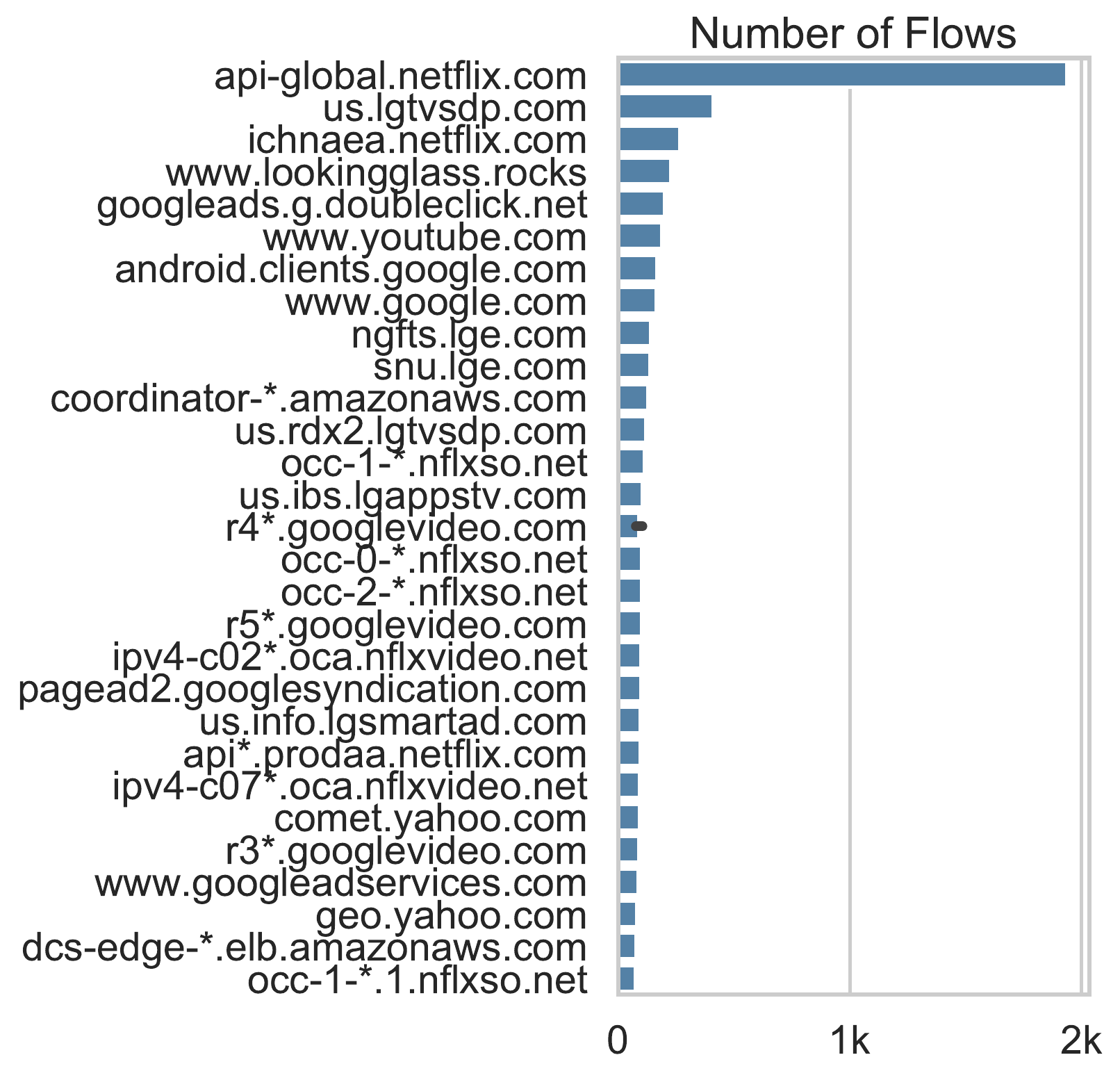}
		\caption{LG}
	\end{subfigure}

	\caption{Continuation of Fig.~\ref{fig:in-the-wild-fig} from Sec.~\ref{sec:in-the-wild}: top-30 fully qualified domain names in terms of number of flows per device for the remaining devices in the in-the-wild dataset.}
	\label{fig:in-the-wild-fig-append}
\end{figure}

\end{document}